\documentclass[fleqn,usenatbib]{mnras}

\usepackage{newtxtext,newtxmath}

\usepackage[T1]{fontenc}

\DeclareRobustCommand{\VAN}[3]{#2}
\let\VANthebibliography\thebibliography
\def\thebibliography{\DeclareRobustCommand{\VAN}[3]{##3}\VANthebibliography}

\usepackage{graphicx}	
\usepackage{amsmath}	

\title[PG Blazar VLA Study]{A Polarimetric Study of 9 PG Quasars with the VLA}

\author[J. Baghel et al.]{
Janhavi Baghel,$^{1}$\thanks{E-mail: jbaghel@ncra.tifr.res.in}
P. Kharb,$^{1}$
Silpa S.,$^{1}$
Luis C. Ho,$^{2, 3}$
C. M. Harrison$^{4}$
\\
$^{1}$National Centre for Radio Astrophysics (NCRA) - Tata Institute of Fundamental Research (TIFR), S. P. Pune University Campus, Ganeshkhind, Pune 411007, India \\
 $^{2}$Kavli Institute for Astronomy and Astrophysics, Peking University, Beijing 100871, China\\
 $^{3}$Department of Astronomy, School of Physics, Peking University, Beijing 100871, China\\
 $^{4}$School of Mathematics, Statistics and Physics, Newcastle University, Newcastle upon Tyne NE1 7RU, UK
}

\begin{document}
\label{firstpage}
\pagerange{\pageref{firstpage}--\pageref{lastpage}}
\maketitle

\begin{abstract}
We present polarization images of 9 radio-loud (RL) quasars from the VLA B-array at 6~GHz. These quasars belong to the Palomar-Green (PG) ``blazar'' sample comprising 16 RL quasars and 8 BL~Lac objects. Extensive polarization is detected in the cores, jets and lobes of all the quasars, with cores primarily displaying magnetic (B-) fields transverse to, and jets displaying fields aligned with the jet direction. Hotspots display either transverse B-fields signifying B-field compression at terminal shocks or more complex structures. The fractional polarization in the cores ranges from 1-10\% and jets/lobes from 10-40\%. Several of the quasars show distorted or hybrid FRI/FRII radio morphologies with indications of restarted AGN activity. We attribute this to the optical/UV selection criteria of the PG sample that remains unbiased at radio frequencies. The in-band spectral indices of the radio cores are relatively flat while they are steep in the hotspots. This is consistent with the polarization structures where the hotspots appear to be locations of jet bends or bow-shocks. We present global properties for the entire PG ``blazar'' sample. We find that jet powers correlate with accretion rates for the quasars; higher accretion rates result in more powerful radio jets. A correlation between the radio core fractional polarization and the 150~MHz total radio luminosity for the 9 quasars studied here may imply that more organized B-fields at the jet bases lead to higher core fractional polarization and to more powerful radio jets.
\end{abstract}


\section{Introduction}
Active Galactic Nuclei (AGN) are highly energetic and luminous non-stellar sources in the centres of galaxies that are powered by an actively accreting supermassive black hole \citep[SMBH;][]{Rees84,Rawlings1991}. Radio-loud AGN (RL AGN) are AGNs with collimated, bipolar, relativistic jets \citep{Blandford1974}, parameterised as having {$\mathrm {R = (S_{5~GHz}/S_{{\it B}-band}) \geq 10}$} where R is the ratio of radio (5~GHz) to optical ({\it B}-band) flux densities \citep{Kellermann1989}. Only around $\mathrm{15-20\%}$ of AGN are radio-loud \citep{Kellermann1989, Urry2003} but these kpc–Mpc scale jetted AGN have a major impact on their host galaxy and surrounding environments, stimulating and limiting the growth of galaxies, star-formation, and heating the circumgalactic medium \citep[][]{Blandford2019}. The different classes of RL AGN are radio galaxies (RGs), quasars and  BL~Lac objects. RGs and quasars are distinguished by the presence of narrow, as opposed to broad (velocity widths of $\sim 1000$~km~s$^{-1}$) optical emission lines respectively \citep[][]{Osterbrock1980, Wills1986}, whereas the optical spectra of BL~Lacs is mostly featureless and dominated by the continuum emission \citep{Stickel1991,Stocke1991}.

The prevailing jet based RL AGN model is intrinsically anisotropic and orientation based RL unification postulates that observational properties of RL AGNs (and in general of Type 1 and 2 AGN) are explained by the different orientation of an opaque dust distribution (torus) relative to an observer and the effects of relativistic beaming \citep[][]{Barthel1989, Antonucci1993, Urry95}. The torus lies in the radio axis equatorial plane, surrounding the central engine and broad emission line region (BLR) obscuring them from view for certain viewing angles, resulting in Type 1 (broad + narrow optical emission lines; broad line RGs and quasars) and Type 2 (narrow line RGs) AGN. Relativistic beaming of jets oriented at small angles to the observer leads to superluminal velocities, one-sided core-jet morphologies, depolarization asymmetry, and high source brightness. However, there are other phenomenological differences in RL AGN unrelated to the viewing angle. A morphological divide in radio jets of RGs was first noted by \citet{Fanaroff1974} and hence they are denoted as Fanaroff \& Riley Type I (FRI) and Type II (FRII). The lower-luminosity FRI RGs exhibit plume-like radio lobes whereas the higher-luminosity FRII RGs exhibit collimated jets with terminal ``hotspots''. A division in total radio power found to occur at $\mathrm{L_{178~MHz} = 2 \times 10^{25}~W~Hz^{-1}~sr^{-1}}$ \citep{Fanaroff1974}.
 
The RL unification posits that quasars are the pole-on counterparts of FRII RGs while BL~Lac objects are the pole-on counterparts of FRI RGs \citep{Barthel1989,Urry95}. Blazars have been historically characterised as AGN with high luminosities, rapid variability, high and variable polarization, superluminal jet motion, and intense non-thermal emission across the electromagnetic spectrum. These properties are understood to be now largely explained by their relativistic jets being oriented at small angles to the line of sight, where both Doppler and projection effects become prominent. In addition, blazar spectral energy distributions (SEDs) exhibit double peaks, attributed to synchrotron radiation and inverse Compton (IC) radiation, respectively \citep[e.g.,][]{Abdo2010}. While quasars are found to have low synchrotron peak frequency $\mathrm{(\nu_{peak} < 10^{14.5}~Hz)}$, BL Lacs are found to span the entire range, with low-, intermediate-, and high-frequency-peaking BL Lacs \citep[LBL, IBL, and HBL;][]{Ghisellini2011}.

The origin of the FR dichotomy in radio power and morphology is still an unresolved question. As is the question of the dramatically different optical spectra and broad-band SEDs in the blazar subclasses, {\it aka} the blazar divide. Various hypotheses have been put forth to explain the differences between the FR radio galaxies as well as the blazar sub-classes. Two popular suggestions include {``intrinsic differences''} between FRIs and FRIIs in terms of jet kinetic power, driven by differences in the SMBH spins \citep[e.g.,][]{Baum1995,Celotti1997,Bhattacharya2016}, or efficiency of accretion \citep[e.g.,][]{Meier1997, Marchesini2004, Ghisellini2010}, and {``extrinsic differences''} in jet-medium interaction that results from differences in the surrounding medium \citep[e.g.,][]{Gopal1996,Urry95, Prestage1988}. 

High-excitation radio galaxies (HERGs) displaying strong optical emission lines are associated with high-rate, radiatively-efficient \citep{Shakura1973} accretion whereas low-excitation radio galaxies (LERGs) with faint or absent lines are linked to low-rate, radiatively-inefficient \citep{Narayan1995} accretion. FRI and FRII classes have been frequently associated with RI and RE accretion states, respectively \citep{Jackson1997}. Almost all FRIs are LERGs and most FRIIs are HERGs. However, there are few known examples of FRI HERGs \citep{Gurkan2021}, and a significant population of FRII LERGs \citep{Hine1979}. This association is also contradicted by the discovery of FRII-like BL Lacs, and FRI-like quasars \citep{Landt2006,Kharb10}. Jet deceleration and subsequent decollimation in FRI jets, which likely occurs on kpc scales  \citep{Bicknell1994,Laing2002} could be controlled by a combination of jet power and environmental density \citep{Ledlow1996}. This suggestion has found support in recent LOFAR observations \citep{Mingo2019,Mingo2022}. 

Jet formation mechanisms, either through the Blandford-Znajek \citep[BZ;][]{BZ1977} or the Blandford-Payne \citep[BP;][]{BP1982} mechanisms, or a combination of both, may also be driving the FR/blazar divide. Several theoretical models of AGN jets predict the generation of helical B-fields propagating outward with the jet plasma \citep{Meier2001,Lyutikov2005,Hawley2015}. Polarization observations are therefore important to understand and differentiate between different jet formation mechanisms. Earlier polarization studies of AGN jets have revealed that fractional polarization up to 40\% are common in radio jets at low radio frequencies \citep{Bridle1984}. The inferred B-field structures \citep[inferred to be perpendicular to the electric vector positions angles, EVPA, for optically thin emission, and parallel to the EVPA for optically thick emission;][]{Pacholczyk70} can be categorised as (i) B-fields predominantly parallel to the jet axis, (ii) B-fields predominantly perpendicular to the jet axis, and, (iii) B-fields perpendicular to the jet axis at the centre of the jet, but parallel near one or both of its edges. More powerful FRII sources tend to have B-fields parallel to the jet axis whereas FRI sources typically display jets with the other two B-field configurations \citep{Willis1981,Bridle1982}. In the jets of most powerful quasars, the inferred B-field direction after correction for Faraday rotation is normally along the jet, often following bends in the jet very closely \citep{Bridle1994}.

Very long baseline interferometry (VLBI) observations of blazars have found differences in magnetic (B-) field structures as well as rotation measures (RM) on parsec-scales \citep{Cawthrone1993}. BL~Lacs tend to have their parsec-scale electric vector position angles (EVPA) parallel to the jet direction whereas RL quasars tend to show a perpendicular relative orientation \citep{Lister2005,Lister2013}. Also, BL~Lacs have been found to show systematically higher parsec-scale RM relative to the quasars \citep{Zavala2005}. However, it is not clear if these differences extend to kpc-scales.

In this paper, we present polarization-sensitive Karl G. Jansky Very Large Array (VLA) images for the jets and lobes of 9 RL quasars to study their kpc-scale B-field structures. The selection criteria for our sample is discussed in Section (2).  We have detailed the radio data reduction and calibration and imaging details in Section (3). Results from our observations are discussed in Section (4) along with a discussion on previous observations of our sources. Global correlations for the entire Palomar Green (PG) ``blazar'' sample have also been discussed in Section (5). We discuss our results and findings in Section (6) and summarize and present our final conclusions in Section (7).

Throughout this paper, we have adopted $\Lambda$CDM cosmology with $\mathrm{H_0 = 73~km~s^{-1} Mpc^{-1}}$, $\mathrm{\Omega_m = 0.3}$ and $\mathrm{\Omega_v = 0.7}$ and used flat $\Lambda$CDM subroutine of {\tt{astropy.cosmology}} subpackage \citep{astropy:2013,astropy:2018}. The spectral index $\alpha$ is defined such that flux density at frequency $\nu$, $S_\nu \propto \nu^\alpha$.

\section{The PG ``Blazar'' Sample}
The Palomar Green UV-excess photographic survey was carried out in the U and B bands with the Palomar 18 inch Schmidt telescope in the late 1970s \citep{Green1986}. The PG UV-excess survey remains the largest complete optically selected survey for unobscured AGN at low redshift and therefore has no radio selection biases. This makes this a good sample to examine the FR dichotomy, which was originally observed in the radio-selected 3C sample. The PG sample is one of the most well studied samples of AGN \citep{Boroson1992,Kellermann1989,Miller1993} having extensive supplementary multi-band data available in the literature, including accurate black hole (BH) masses, accretion rates, galactic properties including star-formation rates (SFR), etc.  \citep[e.g.,][]{Shangguan_2018,Davis2011,Xie2021}. While the PG sample is well studied at optical/UV wavebands, sensitive high-resolution polarimetric radio observations have been lacking for most of the sources. These high-resolution polarimetric data are meant to fill that gap and provide a means to adequately resolve jet structures and to probe the environment through which jets are propagating.

The complete PG sample includes 1715 objects and covers a region of the sky that is 10,714 square degrees. The limiting B-band magnitude for the PG sample is B = 16.1 mag. The sample was 84\% complete in 1986. Only 9\% of the PG sample sources are extragalactic in origin. Quasi-stellar objects (QSOs) constitute 5.4\% of the sample. QSOs include both RL and RQ quasars, with the latter making the vast majority ($>80$\%). Only 4 BL~Lac objects were initially identified in the PG sample. These are: PG0851+203 (OJ287), PG1218+304, PG1418+546, PG1553+113. \citet{Padovani1995} revised this list to 9: apart from the original 4, 3 BL~Lacs were identified by \citet{Fleming1993} : PG1246+586, PG1424+240 and PG1437 +398; and the two identified by \citet{Padovani1995} were PG1101+384 (Mrk421) and PG2254+075. \citet{Padovani1995} also noted that the PG sample may be missing 40\% of BL~Lacs based on its U-B color limit.

\begin{figure}
\centering
\includegraphics[width = 8.5cm,trim=10 20 0 0]{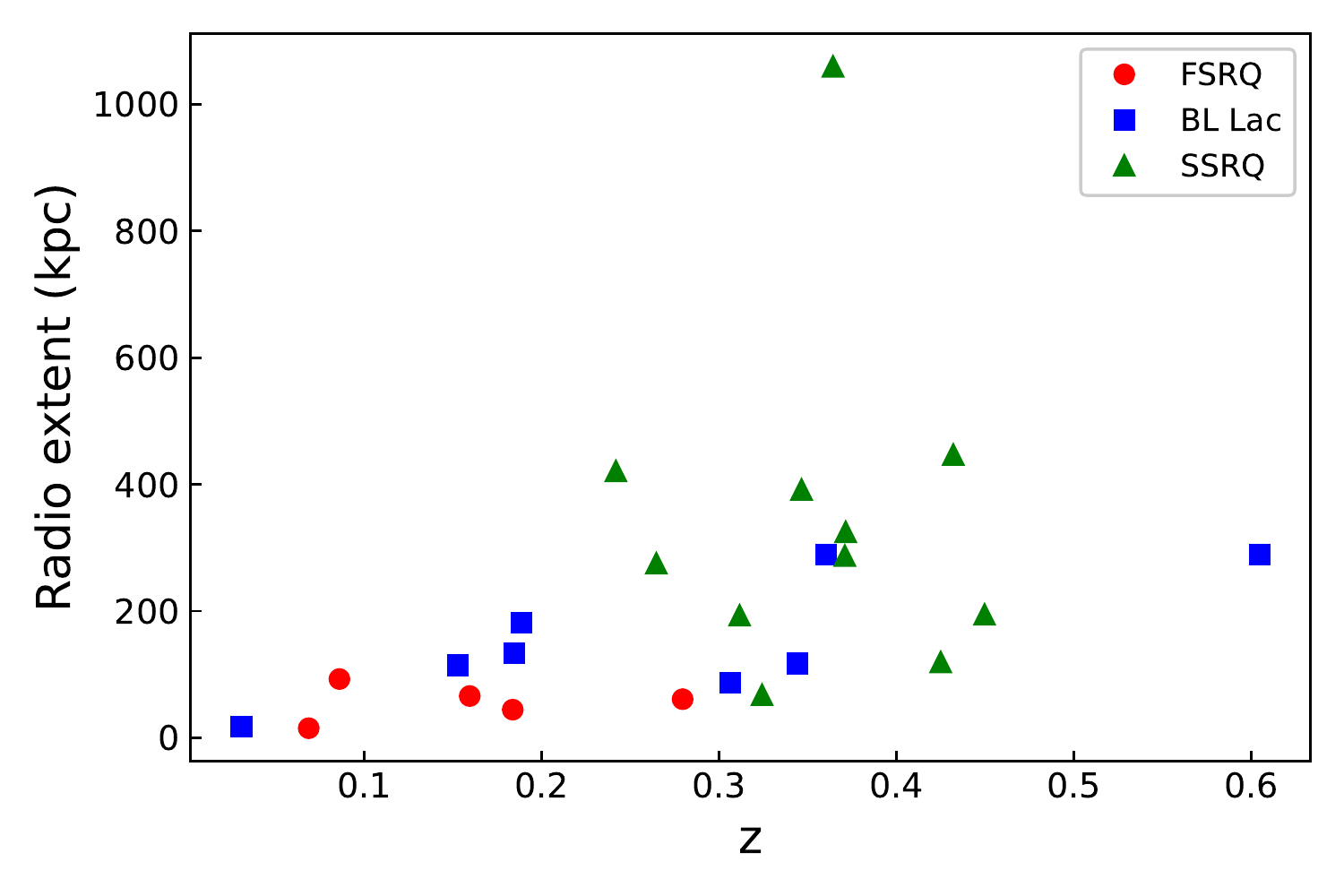}
\caption{\small Distribution of redshifts $z$ w.r.t. the radio extents in kpc for the PG ``blazar'' sample.}
	\label{corr1}
\end{figure}

The PG ``blazar'' sample presented in this paper comprises 16 RL quasars and 8 BL~Lac objects. Historically, blazars include only BL~Lac objects and flat spectrum radio quasars (FSRQs). We have also included steep spectrum radio quasars (SSRQs) in our present study, and loosely referred to the BL~Lacs, FSRQs and SSRQs collectively as ``blazars''. This was chosen to increase the sample size while still including only Type~1 sources, where the jets are inclined relatively close to our lines of sight. Given that the half-opening angle of the dusty obscuring tori is $\leq50\degr$ \citep{Simpson96}, our selection restricts the quasar jet orientations to be $\leq50\degr$ while displaying broad ($\geq1000$~km~s$^{-1}$ velocity widths) emission lines in their optical/UV spectra. Nearly 1/3 of our sample quasars are optically-violently-variable (OVV) quasars. In the paper, when we refer to FSRQs and BL~Lacs exclusively, we call them ``classical'' blazars. Our PG ``blazar'' sample was chosen using the following selection criteria.
\begin{enumerate}
\item Redshift $z<0.5$
\item Projected core-to-lobe extents $\gtrsim15\arcsec$ 
\end{enumerate}
The projected lobe extents were chosen so that the $\sim5\arcsec$ uGMRT Band~4 ($\mathrm{\sim 600~MHz}$) observations can sample the lobes three or more times. This results in the $\sim1\arcsec$ VLA 6~GHz B-array observations probing at least $\sim5$~kpc scale regions at the distance of these sources. 1 BL~Lac object, viz., PG1246+586, was excluded due to the redshift cutoff and 1 PG RL quasar, viz., PG1211+143, was excluded as it was unresolved on kpc-scales \citep{Danehkar2018}. {  In general, Type 1 AGN are good candidates for carrying out polarimetric observations as smaller inclination angles and Doppler boosting effects are conducive to reduced depolarising effects from intervening media as well as radio flux density brightening.} 
Figure~\ref{corr1} presents the range of redshifts and  spatial extents in kpc of the PG ``blazar'' sample. In this paper, we present the VLA images of 9 sample RL quasars which we had already acquired data for from our PG blazar sample\footnote{ New VLA observations have recently been approved for the remaining 15 sources.}. \citet{Silpa2021} have already discussed source PG0007+106 using this data but we mention those results as this source is also part of our sample and we also provide the in-band spectral index image of the same.

\begin{table*}
\centering
\caption{\label{tab:PG}The PG ``Blazar'' Sample}
\begin{tabular}{c|c|c|c|c|c|c|c|c}
\hline
S.No. & Name  
& Other Name  & RA & DEC & Redshift & Radio Extent ($\arcsec$ | kpc) & Type \\
\hline

1 &
{PG0851+203} &
OJ 287 &
08h54m48.87s &
+20d06m30.64s &
0.306501 &
20 | 87 & 
BL Lac \\

2 &
{PG1101+384} &
Mrk 421 &
11h04m27.31s &
+38d12m31.79s &
0.030893 &
30 | 18 & 
BL Lac \\

3 &
{PG1218+304} &
RBS 1100 &
12h21m21.94s &
+30d10m37.16s &
0.184537 &
45 | 134 &
BL Lac \\

4 &
{PG1418+546} &
OQ +530 &
14h19m46.59s &
+54d23m14.78s &
0.152845 &
45 | 115 & 
BL Lac\\

5 &
{PG1424+240} &
OQ +240&
14h27m00.39s &
+23d48m00.03s &
0.160680$^\ast$ &
45 | 290 & 
BL Lac\\

6 &
{PG1437+398} &
RBS 1414 &
14h39m17.47s &
+39d32m42.80s &
0.344153 &
{ 25 | 117 } & 
BL Lac\\

7 &
{PG1553+113} &
RBS 1538 &
15h55m43.04s &
+11d11m24.36s &
0.360365 &
60 | 290 & 
BL Lac\\

8 &
{PG 2254+075} &
OY +091 &
22h57m17.30s &
+07d43m12.30s &
0.188765 &
60 | 181 & 
BL Lac\\

9 &
{PG0007+106} &
Mrk 1501 &
00h10m31.00s &
+10d58m29.50s &
0.086032 &
60 | 93 & 
FSRQ \\

10 &
{PG1226+023} &
3C 273 &
12h29m06.69s &
+02d03m08.59s &
0.159492 &
25 | 66 &
FSRQ \\

11 &
{PG1302$-$102} &
RBS 1212&
13h05m33.01s &
-10d33m19.43s &
0.27949 &
15 | 61 & 
FSRQ \\

12 &
{PG1309+355} &
Ton 1565 &
13h12m17.75s &
+35d15m21.08s &
0.183734 &
{ 15 | 44 }& 
FSRQ\\

13 &
{PG2209+184} &
II Zw 171 &
22h11m53.88s &
+18d41m49.85s &
0.06873 &
{ 12 | 15 }&
FSRQ\\

14 &
{PG0003+158} &
4C +15.01 &
00h05m59.23s &
+16d09m49.02s &
0.449711 &
35|194 &
SSRQ
\\

15 &
{PG1004+130} &
4C +13.41 &
10h07m26.09s &
+12d48m56.18s &
0.24186  &
115|421 &
SSRQ
\\

16 &
{PG1048$-$090} &
3C 246 &
10h51m29.91s &
-09d18m10.19s &
0.346528 &
83|392   &
SSRQ
\\

17 &
{PG1100+772} &
3C 249.1 &
11h04m13.86s &
+76d58m58.19s &
0.31163 &
44|193 & 
SSRQ
\\

18 &
{PG1103$-$006} &
4C -00.43 &
11h06m31.77s &
-00d52m52.38s &
0.424968 &
22|119 &
SSRQ 
\\

19 &
{PG1425+267} &
Ton 202 &
14h27m35.60s &
+26d32m14.54s &
0.364262  &
218|1060 &
SSRQ
\\

20 &
{PG1512+370} &
4C +37.43 &
15h14m43.06s &
+36d50m50.35s &
0.370922 &
58|287 &
SSRQ
\\

21 &
{PG1545+210} &
3C 323.1 &
15h47m43.53s &
+20d52m16.61s &
0.264659  &
70|275  &
SSRQ 
\\

22 &
{PG1704+608} &
3C 351 &
17h04m41.37s &
+60d44m30.52s &
0.371411 &
66|325 &
SSRQ 
\\

23 &
{PG2251+113} &
4C +11.72 &
22h54m10.42s &
+11d36m38.74s &
0.32427 &
15 | 68 &
SSRQ 
\\

24 &
{PG2308+098} &
4C +09.72 &
23h11m17.75s &
+10d08m15.75s &
0.432064 &
83|447 &
SSRQ
\\
\hline
\multicolumn{8}{l}{Note. Column (1): Serial Number. Column (2): PG names of sources. Column (3): Other common names of sources. Column (4): Right Ascension.}\\
\multicolumn{8}{l}{Column (5): Declination. Column (6): Redshift$^\dagger$. Column (7): Radio extents in arcsec and kpc. Column (9): Type of blazar}\\
\multicolumn{8}{l}{$^\ast$ $z=0.60468$ has been suggested by \citet{Paiano17}. We have used this updated value in our calculations}\\
\multicolumn{8}{l}{$^\dagger$ All redshift values reported are corrected to the reference frame defined by the 3K CMB on NASA NED\footnote{The NASA/IPAC Extragalactic Database (NED)
is operated by the Jet Propulsion Laboratory, California Institute of Technology,
under contract with the National Aeronautics and Space Administration.}.}\\ 
\multicolumn{8}{l}{Radio extents derived from \citet{Miller1993} at 5~GHz or from 1.4~GHz VLA FIRST / NVSS images for sources unresolved at 5~GHz.}
\end{tabular}
\end{table*}

\section{Radio Data: Reduction and Analysis}
{The radio data were acquired using the VLA in the C-band (6~GHz) B-array configuration from 2020, August 10 to 2020, October 15 (Project ID: 20A-182) with a resolution of 1.1$\arcsec$. 16 spectral windows with 64 channels each were chosen to span the frequency range of 4.5 - 6.6~GHz. The average time on source was around 10~min. Polarization calibrators 3C286 and 3C138 were used. For more details see table~\ref{tab1}. We used the CASA calibration pipeline for VLA data reduction for the initial calibration and flagging. This was followed by manual polarization calibration.

First, the model of a polarized calibrator was set manually using the task \texttt{SETJY} in CASA. For this, we provided the model parameters such as the reference frequency, Stokes I flux density at the reference frequency, the spectral index and the coefficients of the Taylor expansion of fractional polarization and polarization angle as a function of frequency about the reference frequency. The coefficients were estimated by fitting a first-order polynomial to the values of fractional polarization and polarization angle as function of frequency, which were obtained from the NRAO VLA observing guide\footnote{ https://science.nrao.edu/facilities/vla/docs/manuals/obsguide/modes/pol}. Stokes I flux density values were calculated from the \citet{Perley2017} scale and used to estimate the spectral index by fitting and provided the coefficients for Stokes I flux density at reference frequency, the spectral index (alpha), and curvature (beta).

\begin{table*}
\centering
\caption{\label{tab:PG2}The PG ``Blazar'' Sample Properties}
\begin{tabular}{c|c|c|c|c|c|c|c|c|c|c}
\hline
S.No. & Name & 
Type & $\mathrm{\log_{10} (M_{BH} / M_{\sun})}$ & Ref & R & ${\log_{10} \dot{M} \mathrm{(M_{\sun}/yr)}}$ 
&  Ref &  $\mathrm{\log_{10} SFR(M_{\sun}/yr)}$ & Ref & $\bar{Q}$ ($10^{42}$erg/s) 
\\ \hline

1 &
{PG0851+203} &
BL Lac &
8.5 & 1 & ... &
...&...&...
& ... & 210 
\\

2 &
{PG1101+384} &
BL Lac &
8.23 & 1 & ... &
... & ... & ...
& ... & 6.58 
\\

3 &
{PG1218+304} &
BL Lac &
8.47 & 1 & ... &
... & ... & ... 
& ... & 9.23 
\\

4 &
{PG1418+546} &
BL Lac &
8.74 & 1 & ... & ... & ... 
& ... & ... & 61.80 
\\

5 &
{PG1424+240} &
BL Lac &
6.42 & 1 & ... &  ... 
& ... & ...& ... & 988 
\\

6 &
{PG1437+398} &
BL Lac &
8.95 & 1 & ... & ... 
& ... & ...& ... & 165 
\\

7 &
{PG1553+113} &
BL Lac &
7.25 & 1 & ... & ... 
& ... & ...& ... & 120 
\\

8 &
{PG 2254+075} &
BL Lac &
8.85 & 1 & ... & ... 
& ... & ... & ... & 43 
\\

9 &
{PG0007+106} &
FSRQ &
8.87 & 2 & 197 &
-0.42 
& 3 & 0.92 & 5 & 7.01 
\\

10 &
{PG1226+023} &
FSRQ &
9.18 & 2 & 1138 &
1.18 
& 3 &  1.56 & 6 & 4850 
\\

11 &
{PG1302$-$102} &
FSRQ &
9.05 & 2 & 187 &
0.92 
& 3 & 1.82 & 5 & 161 
\\

12 &
{PG1309+355} &
FSRQ &
8.48 & 2 & 18 &
0.37 
& 3 & 1.06 & 5 & 9.92 
\\

13 &
{PG2209+184} &
FSRQ &
8.89 & 2 & 141 &
-0.98 
& 3 & 0.46 & 5 & 1.50 
\\

14 &
{PG0003+158} &
SSRQ &
9.45 & 2 & 175 &
0.79 
& 3 & $<1.25$ & 5 & 2380 
\\

15 &
{PG1004+130} &
SSRQ &
9.43 & 2 & 228 &
-0.37 
& 4 & 1.42 & 5 & 847 
\\

16 &
{PG1048$-$090} &
SSRQ &
9.37 & 2 & 377 &
0.3 
& 3 & $<1.15$ & 5 & 2690 
\\

17 &
{PG1100+772} &
SSRQ &
9.44 & 2 & 322 &
0.29 
& 3 & 1.83 & 5 & 3540 
\\

18 &
{PG1103$-$006} &
SSRQ & 
9.49 & 2 & 272 &
0.21 
& 3 & $<1.15$ & 5 & 2250 
\\

19 &
{PG1425+267} &
SSRQ &
9.9 & 2 & 53.6 &
0.07 
& 3 & 1.79 & 5 & 519 
\\

20 &
{PG1512+370} &
SSRQ &
9.53 & 2 & 190 &
0.2 
& 3 & 1.19 & 5 & 1920 
\\

21 &
{PG1545+210} &
SSRQ & 
9.47 & 2 & 418 &
0.01 
& 3 & $<1.15$ & 5 & 1930 
\\

22 &
{PG1704+608} &
SSRQ & 
9.55 & 2 & 645 &
0.38 
& 3 & 1.94 & 5 & 4640 
\\

23 &
{PG2251+113} &
SSRQ & 
9.15 & 2 & 365 &
0.66 
& 3 & 0.46 & 5 & 1760 
\\

24 &
{PG2308+098} &
SSRQ &
9.76 & 2 & 188 &
0.22 
& 3 & $<1.25$ & 5 & 2000 
\\\hline
\multicolumn{11}{l}{Note. Column (1): Serial Number. Column (2): PG names. Column (3): Blazar type. Column (4): Black hole masses. Column (5): References for black hole masses.}\\
\multicolumn{11}{l}{Column (6): Radio loudness parameter from \citet{Kellermann1994}. Column (7): Accretion rates. Column (8): References for accretion rates. Column (9): Star  }\\
\multicolumn{11}{l}{Formation Rates (SFR). Column (10): References for SFR. Column (11): Jet Power.}\\
\multicolumn{11}{l}{References- 1: \citet{Wu_2009}, 2: \citet{Shangguan_2018}, 3: \citet{Davis2011}, 4: \citet{Luo2013}, 5: \citet{Xie2021}, 6: \citet{Westhues2016}}\\
\end{tabular}
\end{table*}

\begin{table*}
 \caption{VLA Observational Details of the 9 PG quasars}
 \label{tab1}
 \begin{tabular}{|c|c|c|c|c|c|c|}
  \hline
 Source & Observation Date & Time on source & Flux Calibrator & Phase Calibrator & Leakage Calibrator & Polarization Angle Calibrator \\
  \hline
  PG0003+158 & 20-Aug-2020 & 10m42s & 3C138 & J0010+1724 & 3C84  & 3C138 \\
  PG0007+106 & 20-Aug-2020 & 10m54s & 3C138 & J0010+1724 & 3C84  & 3C138 \\
  PG1004+130 & 25-Aug-2020 & 10m42s & 3C138 & J1016+2037 & 3C84  & 3C138 \\
  PG1048-090 & 27-Sep-2020 & 10m51s & 3C138 & J1130-1449 & 3C84  & 3C138 \\
  PG1100+772 & 22-Aug-2020 & 10m24s & 3C138 & J1044+8054 & 3C84  & 3C138 \\
  PG1103-006 & 22-Aug-2020 & 10m30s & 3C138 & J1058+0133 & 3C84  & 3C138 \\
  PG1226+023 & 15-Oct-2020 & 10m36s & 3C286 & J1150-0023 & 3C286 & 3C286 \\
  PG1309+355 & 12-Oct-2020 & 10m30s & 3C286 & J1310+3220 & 3C286 & 3C286 \\
  PG1704+608 & 10-Aug-2020 & 07m06s & 3C286 & J1740+5211 & 3C286 & 3C286 \\
  \hline
  \multicolumn{7}{l}{Note. Column(1): PG source name. Column(2): Observation date. Column(3): Time on source. Column(4): Flux calibrator.}\\
  \multicolumn{7}{l}{ Column(5): Phase calibrator. Column(6): Leakage calibrator Column(7): Polarization angle calibrator}
 \end{tabular}
\end{table*}

Polarization calibration was carried out in three steps: (i) the cross-hand (RL, LR) delays, were solved using a polarized calibrator with strong cross-polarization (either 3C138 or 3C286). This was carried out using the task \texttt{GAINCAL} with gaintype = \texttt{KCROSS} in CASA. (ii) the instrumental polarization (i.e, the frequency dependent leakage terms or ‘D-terms’), was solved using a polarized calibrator with good parallactic angle coverage. We used the task \texttt{POLCAL} in CASA to solve for instrumental polarization with with poltype = Df + QU while using the polarized calibrators (either 3C138 or 3C286) and poltype = Df  while using the { un}polarized calibrator (3C84). The average value of the D-term amplitude turned out to be $\approx$ 7\% (iii) the frequency-dependent polarization angle was solved using a polarized calibrator with known electric vector position angle, EVPA (either 3C138 or 3C286). This was carried out using the task \texttt{POLCAL} in CASA with poltype = Xf.

The calibration solutions were then applied to the multi-source dataset. The CASA task \texttt{SPLIT} was used to extract the calibrated visibility data for the sources from the multi-source dataset while averaging the spectral channels such that the bandwidth (BW) smearing effects were negligible. The total intensity or the Stokes I image of the sources created using the multiterm-multifrequency synthesis \citep[MT-MFS;][]{MTMFS2011} algorithm of \texttt{TCLEAN} task in CASA. {  Three rounds of phase-only self-calibration followed by one round of amplitude and phase self-calibration were carried out for almost all the datasets, except those for PG1704+608 and PG1309+355 for which only two rounds of phase-only self-calibration were carried out.} { Images were made using natural weighting with {\tt robust=+0.5} in {\tt CASA}. For PG1048$-$090, we created the images using robust=+2 in order to reduce the Y-shaped deconvolution errors that showed up around the hotspots.} Stokes Q and U images were created from the final self-calibrated visibility data.

Linear polarized intensity $\mathrm{P = \sqrt{Q^2 + U^2}}$ and the EVPA $\chi = 0.5~\mathrm{{tan}^{-1} (U/Q)}$ images were obtained by combining the Stokes Q and U images using the AIPS\footnote{Astronomical Image Processing System; \citet{Wells1985}} task \texttt{COMB} with opcode = \texttt{POLC} (which corrects for Ricean bias) and \texttt{POLA} respectively. Regions with intensity values less than 3 times the rms noise in the $P$ image and with values greater than $10^{\circ}$ error in the $\chi$ image were blanked using \texttt{COMB}. Fractional polarization $\mathrm{FP=P/I}$ images were obtained in \texttt{COMB} with opcode = \texttt{DIV}. Regions with fractional polarization errors $\gtrsim$10\% were blanked in the image. While imaging with the MT-MFS algorithm of the \texttt{TCLEAN} task in CASA, we used two Taylor terms to model the frequency dependence of the sky emission by setting the parameter nterms = 2. This produced in-band spectral index images and spectral index noise images
for individual sources. We blanked the pixels with spectral index errors greater than 0.3.  

The average r.m.s. noise in the Stokes I images is $3 \times 10^{-5}$~mJy~beam$^{-1}$ (except for the outlier PG1226+023 or 3C273). The results from these images are presented in Table \ref{tab4}}. We note that PG1048$-$090 was imaged with a circular convolved beam of size of $3.17\arcsec$ to minimise the imaging artefacts, which were difficult to get rid of completely with different weighting schemes. Flux density values reported in the paper were obtained using the Gaussian-fitting AIPS task \texttt{JMFIT} for compact components like the core, and AIPS verb \texttt{TVSTAT} for extended emission. The FP and spectral index values noted are the mean values over the noted region. The rms noise values were obtained using AIPS tasks \texttt{TVWIN} and \texttt{IMSTAT}. AIPS procedure \texttt{TVDIST} was used to obtain spatial extents.

{  
We note that Faraday rotation effects are not expected to be significant at 5 GHz in the kpc-scale observations of jets and lobes \citep[e.g.,][]{Saikia1987,Pudritz2012}. For several of the sample sources, viz., PG1004+130, PG1048$-$090, PG1100+772, and PG1226+023, the average integrated kpc-scale RM values are, respectively, $-16\pm5$, $+0\pm3$, $+29\pm1$, $+2.0\pm 0.2$~rad~m$^{-2}$ \citep{1981Simard}. Integrated RM $\lesssim 50$~rad~m$^{-2}$ result in rotation of $\lesssim 10\degr$ at 5~GHz. RM values can indeed be higher than the integrated RM values in smaller local regions, which can influence the local inferred B-field directions \citep[e.g.,][]{McKean2016,Silpa2021B}. However, the absence of strong Faraday effects is supported by the observed polarization in the jets in these sources which tend to lie either parallel or perpendicular to the local jet direction, with the inferred B field directions being perpendicular to the direction of the EVPA vectors for optically-thin regions of emission; such orientations could be indicative of organised B-field structures \citep[see][]{Pudritz2012}. We had estimated the bandwidth depolarization in 5~GHz VLA data by comparing the ratios of polarized flux density of individual sub-bands (of 512~MHz BW each) to that of the full band (of 2 GHz BW); this turned out to be around a factor of 0.5 \citep{Silpa2022}.}

\begin{table*}
 \caption{Observational Results}
 \label{tab4}
 \begin{tabular}{|c|c|c|c|c|c|c|c|c|}
  \hline
  Source & $I_{rms}$(Jy/beam) & $P_{rms}$(Jy/beam) & Region && P~(Jy)& I~(Jy) & FP~(\%) & $\alpha$ \\
  \hline
  PG0003+158 & 2.752E-05 & 2.390E-06 & Core && (3.6 $\pm$ 0.8) E-04 & 1.65E-01 & { 8 $\pm$ 2\%$^\dagger$}  & -0.13$\pm$0.13 \\
  &&& Hotspot region & NW & (1.8$\pm$ 0.2) E-03 & 3.3264E-02 & 9 $\pm$ 2\% & -1.0$\pm$0.2 \\
  &&& & SE & (4.0 $\pm$ 0.2) E-03 & 7.3481E-02 & 11 $\pm$ 2\% & -1.2$\pm$0.3 \\
  &&& Lobes & NW & (3.5 $\pm$ 0.8) E-03 & 2.8033E-02 & 23 $\pm$ 5\% & -1.2$\pm$0.4\\
  &&&& SE & (4.4 $\pm$ 0.8) E-03 & 4.5004E-02 & 17 $\pm$ 4\% & -1.5$\pm$0.6\\
  &&& SE jet knot && (1.04 $\pm$ 0.05)E-03 & 6.3260E-03 & 17.7 $\pm$ 0.8\% & -1.3$\pm$0.1 \\
  \hline
  PG0007+106 & 1.559E-05 & 2.770E-06 & Core && (1.2 $\pm$ 0.3) E-04 & 1.26E-01 & 0.8 $\pm$ 0.3\% & 0.12$\pm$0.03\\
  &&& Lobes & W & (2.3 $\pm$ 0.5) E-04 & 4.18E-03 & 16 $\pm$ 4\% & -1.0$\pm$0.3 \\
  \hline
  PG1004+130 & 2.641E-05 & 4.769E-06 & Core && (2.7 $\pm$ 0.8) E-05 & 2.90E-02 & {  3 $\pm$ 1\%$^*$} & -0.3 $\pm$ 0.2\\
  &&& Lobes & NW & (2.9$\pm$ 0.3) E-02 & 1.2465E-01 & 28 $\pm$ 4\% & -2.0 $\pm$ 0.2 \\
  &&&& SE & (1.3 $\pm$ 0.3) E-02 & 4.9203E-02 & 31 $\pm$ 9\% & -2 $\pm$ 1\\
  &&& SE jet && (3.0 $\pm$ 0.5) E-03 & 1.5625E-02 & 24 $\pm$ 5 \% & -2.4 $\pm$ 0.5\\
  &&& NW jet knot { /hotspot} && (1.6 $\pm$ 0.5) E-04 & 1.8104E-03 & 26 $\pm$ 8\% & -1.8 $\pm$ 0.7\\
  \hline
  PG1048-090 & {  9.089E-04 }& { 5.375E-05 } & Core && { (8 $\pm$ 2) E-05 }&{  4.9798E-02 }&{  2.1 $\pm$ 0.4 \%  }&{  0.2 $\pm$ 0.4 }\\
  &&& Hotspot region & NW & { (1.4 $\pm$ 0.1) E-02 }&{   2.4785E-01 }&{   10 $\pm$ 2 \% }&{  -0.64 $\pm$ 0.13 }\\
  &&& & SE & { (3.1 $\pm$ 0.1) E-02}  &{  2.8795E-01 }&{  14 $\pm$ 2\%} &{  -1.2 $\pm$ 0.1}\\
   \hline
  PG1100+772 & 3.292E-05  & 4.032E-06 & Core && (1.17 $\pm$ 0.05) E-03 & 1.17E-01 & 2.7 $\pm$ 0.4\% & 0.55$\pm$0.006 \\
  &&& Hotspot region & W & (1.751 $\pm$ 0.009) E-02 & 2.1315E-01 & 10.48 $\pm$ 0.08\% & -1.051$\pm$0.003\\
  &&& & E & (1.314 $\pm$ 0.007) E-02 & 1.1189E-01 & 14.3 $\pm$ 0.7\% & -0.938$\pm$0.004 \\
  &&& Lobes & W & (3.5 $\pm$ 0.2) E-02 & 1.9166E-01 & 23 $\pm$ 2\% & -0.86$\pm$0.07 \\
  &&& & E & (1.8 $\pm$ 0.3) E-02 & 8.5160E-02 & 25 $\pm$ 4\% & -1.2$\pm$0.2\\
  &&& Jet && (6.4 $\pm$ 0.3) E-03 & 5.7812E-02 & 15 $\pm$ 4\% & -0.86$\pm$0.07\\
  &&& Ridge & E & (1.05 $\pm$ 0.07) E-02 & 4.173E-02 & 23 $\pm$ 3\% & -0.68$\pm$0.07 \\
  \hline
  PG1103-006 &2.511E-05 & 2.511E-05 & Core && (2.08 $\pm$ 0.07) E-03 & 1.37E-01 & 5.8 $\pm$ 0.3 \% & -0.319$\pm$0.004\\
  &&& Hotspot region & NW & (2.18 $\pm$ 0.08) E-03 &  3.1218E-02 &13 $\pm$ 1 \% & -0.76$\pm$0.03\\
  &&& & SE & (1.11 $\pm$ 0.02) E-02 & 8.8719E-02 & 18.5 $\pm$ 0.8\%& -0.93$\pm$0.02\\
  &&& Lobes & NW & (1.47 $\pm$ 0.07) E-02&  1.3515E-01 & 19 $\pm$ 2 \% & -0.7$\pm$0.3\\
  &&& & SE & (7.2 $\pm$ 0.7) E-03 & 1.2988E-01 & 20 $\pm$ 2 \% & -0.6$\pm$0.2\\
  &&& & SW & (2.4 $\pm$ 0.5) E-03 & 1.3785E-02 & 25 $\pm$ 6\% & -1.9$\pm$0.2\\
  &&& & NE & (9 $\pm$ 2) E-04 & 2.7121E-03 & 20 $\pm$ 5\%& -0.9$\pm$0.3\\
  &&& NW jet && (7.6 $\pm$ 0.2) E-03 & 1.1045E-01 & 10.2 $\pm$ 0.5 \% & -0.806$\pm$0.008\\
  \hline
  PG1226+023 & 2.237E-03 & 2.439E-04 & Core && (2.128 $\pm$ 0.009)E+00 & 2.16E+01 & 11 $\pm$ 2\% & -0.327 $\pm$ 0.004\\
  &&& Hotspot region & SW & (5.6 $\pm$ 0.2)E-01 & 5.19E+00 & 15 $\pm$ 2\% & -0.902 $\pm$ 0.03\\
  \hline
  PG1309+355 & 8.706E-05 & 9.592E-06 & Core && (1.3 $\pm$ 0.2)E-04 & 4.69E-02 & 0.37 $\pm$ 0.07\% & 0.17 $\pm$ 0.06\\
  \hline
   PG1704+608 & 4.671E-05 & 1.014E-05 & Core && (1.36 $\pm$ 0.08) E-03 & 2.12E-02 & 8.2 $\pm$ 0.9\% &-0.8$\pm$0.7\\
  &&& Hotspot region & NE (up) & (4.54 $\pm$ 0.02) E-02 & 5.2227E-01 &11.5 $\pm$ 0.4\% &-0.96$\pm$0.02\\
  &&&& NE (lw) & (3.75 $\pm$ 0.02) E-02 & 2.1331E-01 & 15.4 $\pm$ 0.9\%& -1.06$\pm$0.04\\
  &&&& SW & (1.8 $\pm$ 0.2) E-03 & 9.6707E-03 & 16 $\pm$ 2\% & -2.0$\pm$0.2\\
  &&& Lobes & NW & (2.0 $\pm$ 0.5) E-03 & 2.3813E-02 & 23 $\pm$ 7\% & -0.2$\pm$0.7 \\
  &&&& NE & (1.4 $\pm$ 0.1) E-02 & 1.6121E-01 & 13 $\pm$ 2\% & -1.2$\pm$0.5 \\
  &&&& SW & (2.2 $\pm$ 0.7) E-03 & 4.1280E-02 & 14$\pm$ 3\%&-1.9$\pm$0.5\\
  \hline
  \multicolumn{9}{l}{Note. Column (1): PG source name. Column (2): R.M.S noise in Stokes I (total intensity) image. Column (3):  R.M.S noise in polarized intensity image.  Column (4):}\\
  \multicolumn{9}{l}{ Region of the source and location. Column (5): Polarized flux density. Column (6): Total flux density. Column (7): Fractional Polarization. Column (8): Spectral index} \\
  \multicolumn{9}{l}{{ $^\dagger$ FP at intensity peak position is 0.9 $\pm$ 0.3\%. $^*$ FP at intensity peak position is 0.27 $\pm$ 0.09\%. }}\\
 \end{tabular}
\end{table*}

\section{Results}
We have presented the total intensity and polarization images, as well as the in-band spectral index images in Figures~\ref{fig1}-\ref{fig9}. The quantitative results have been noted in Table \ref{tab4} where we have adopted the definitions of a jets, knots and hotspots as detailed by \citet{Bridle1984}. We find extensive polarization in the cores, jets, lobes and hotspots of all the quasars, with the exception of PG1309+355 which remains largely unresolved but shows core polarization (see Table~\ref{tab4}). The radio morphology shows distortions and complexities. The spectral indices are typically flat/inverted for the radio cores, with the sole exception of PG1704+608, but steep in the jets and hotspots (Table~\ref{tab4}). Below we describe these details for the individual sources.

\subsection{Notes on individual sources}

\subsubsection{PG0003+158} 
\begin{figure*}
\centering
\includegraphics[width=15.5cm,trim=0 0 0 0]{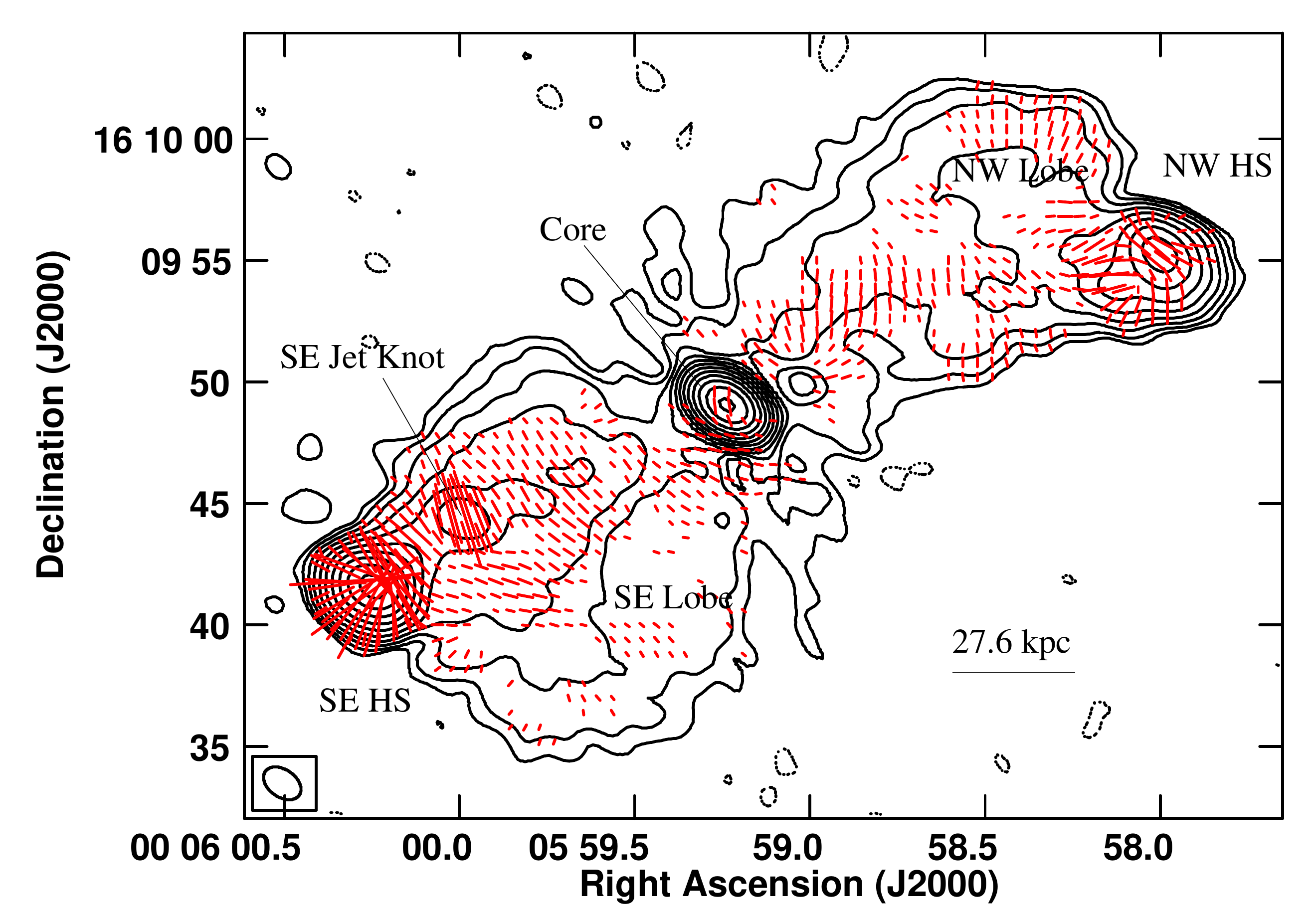}
\includegraphics[width=15.5cm]{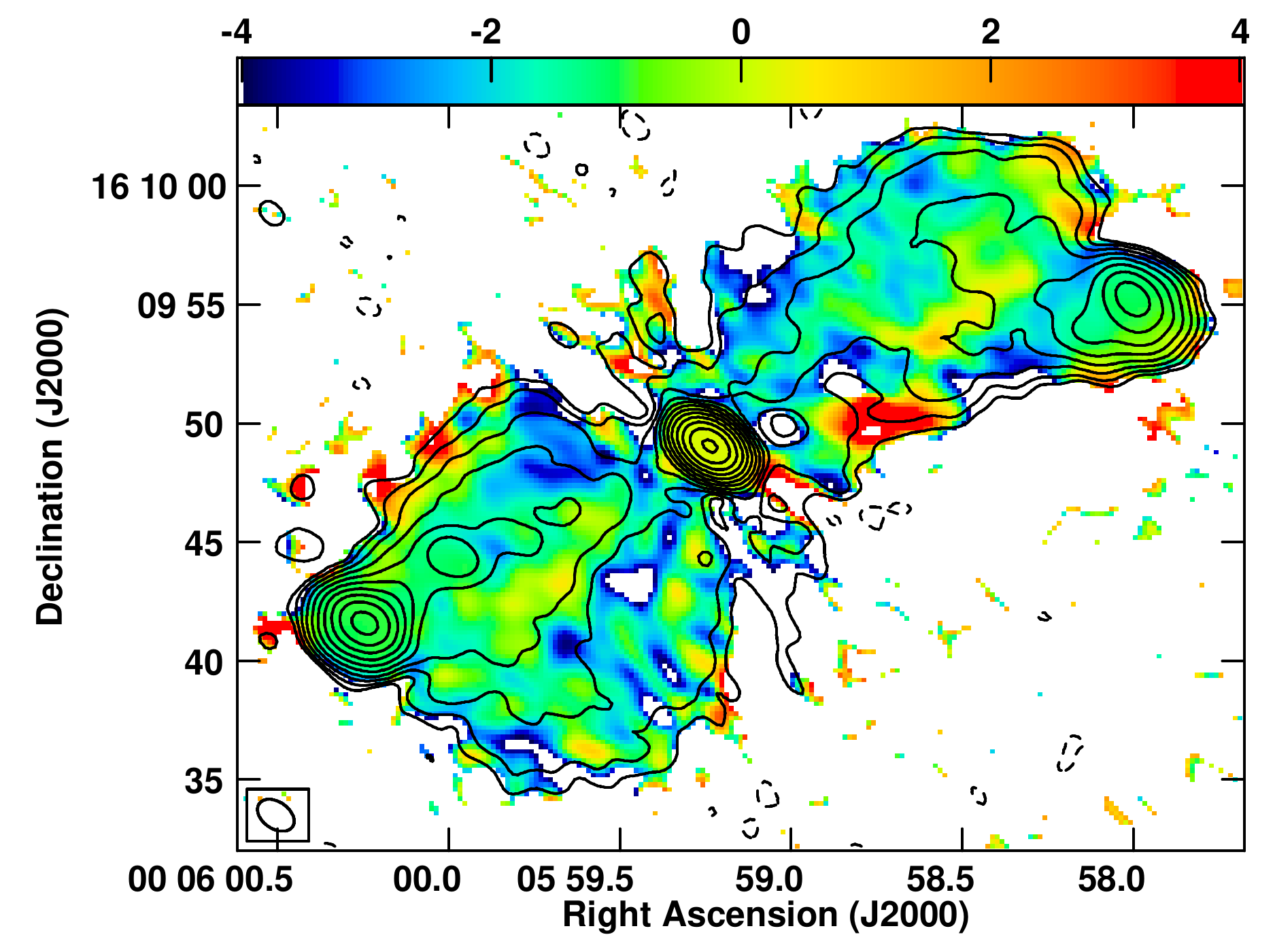}
\caption{\small VLA 6 GHz contour image of quasar PG0003+158 (4C 15.01) superimposed with red (top) polarized intensity vectors and (bottom) in-band spectral index image. The inferred B-fields are perpendicular to the polarization vectors assuming optically thin emission. The beam is $1.71\arcsec \times 1.13\arcsec$ with a PA of $55.18^\circ$. The peak surface brightness, $I_P$ is 161.0~mJy~beam$^{-1}$. The contour levels in percentage of the peak surface brightness $I_P$ are $(-0.045,~0.045,~0.09,~0.18,~0.35,~0.7,~1.4,~2.8,~5.6,~11.25,~22.5,~45,~90)$~Jy~beam$^{-1}$. The length of the EVPA vectors is proportional to polarized intensity with $5\arcsec$ corresponding to 1.25~mJy~beam$^{-1}$.}
\label{fig1}
\vspace{-31.7106pt}
\end{figure*}

This SSRQ ({\it aka} 4C+15.01) was imaged with the VLA A-array at 5~GHz by \citet{Miller1993} who detected only the bright and compact regions (i.e., core and hotspots) of this large double radio source. Our image detects the diffuse lobe emission as well and shows extensive polarization in its core, lobes and hostpots (Figure~\ref{fig1}). The inferred B-field is transverse to the jet direction for the optically thick radio core. The southern jet knot has an inferred B-field morphology inconsistent with a shocked region; the B-field vectors are aligned with the jet flow rather than perpendicular. This jet knot could be a site for particle re-acceleration/energizing which should show a relatively flat spectral index without B-field enchancement. However, the spectral index of the jet knot is $-1.3\pm0.1$, making it steeper than the $\alpha$ of the hotspots ($-1.2\pm0.3$ for the southern hotspot and $-1.0\pm0.2$ for the northern hotspot), suggesting this to be a portion of the jet itself. This is then consistent with the findings of \citet{Bridle1994} who found that jet knots that are elongated in directions close to that of the jet, the EVPA vectors tend to be orthogonal to the jet axis. The inferred B-fields in the lobes are largely aligned with the local jet direction assuming optically thin emission. A jet region entering the southern hotspot is highly polarized, exhibiting parallel inferred B-fields. The inferred B-fields are aligned with the edges of the hotspot regions. Recent VLBI results from \citet{Wang2022} indicate that the parsec-scale approaching jet is towards the south-east but offset from the straight line connecting the core to the southern hotspot, indicating jet curvature.

\subsubsection{PG0007+106} 
\begin{figure*}
\centering
\includegraphics[width=16.75cm,trim=0 0 0 0]{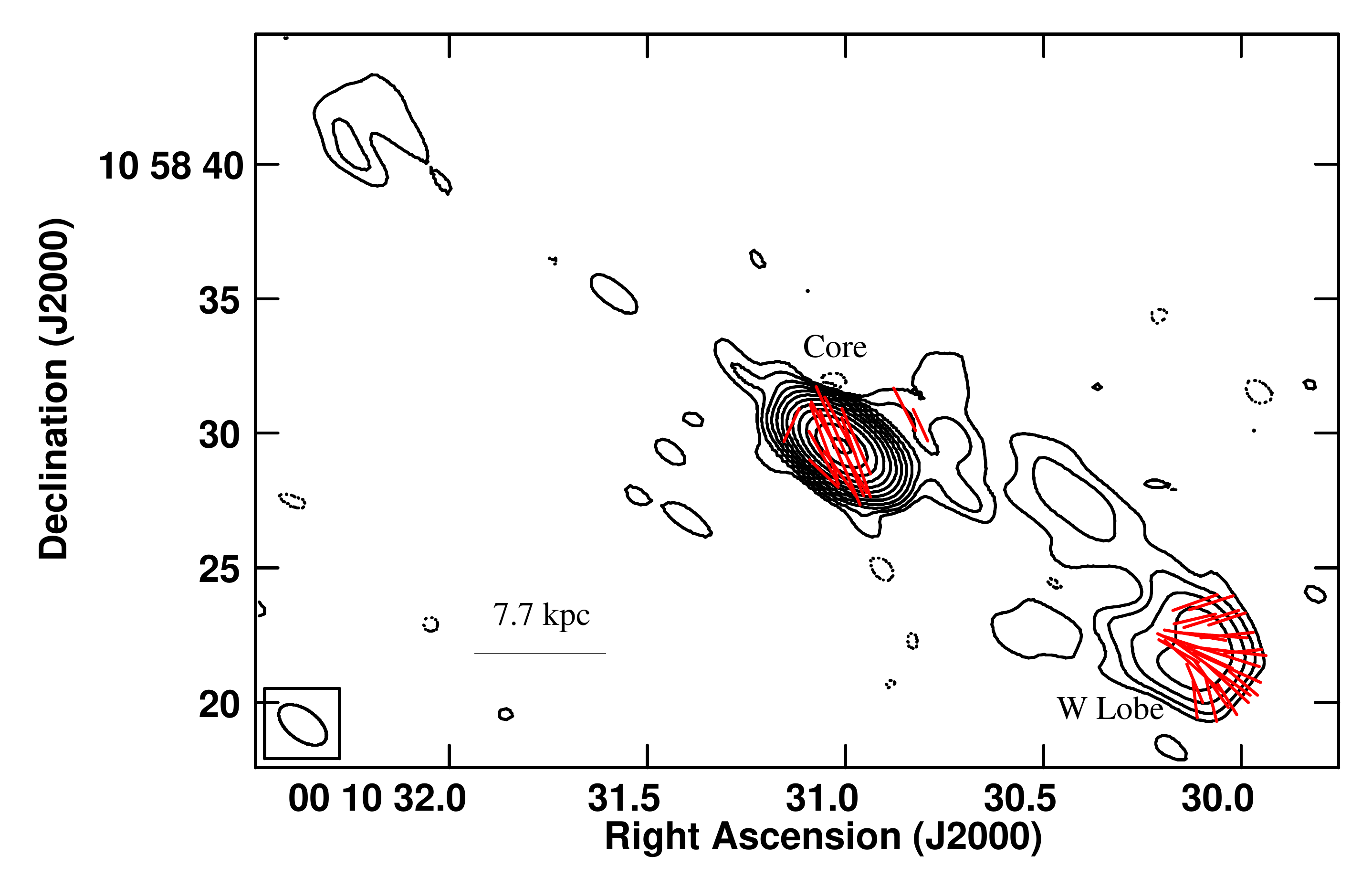}
\includegraphics[width=16.75cm]{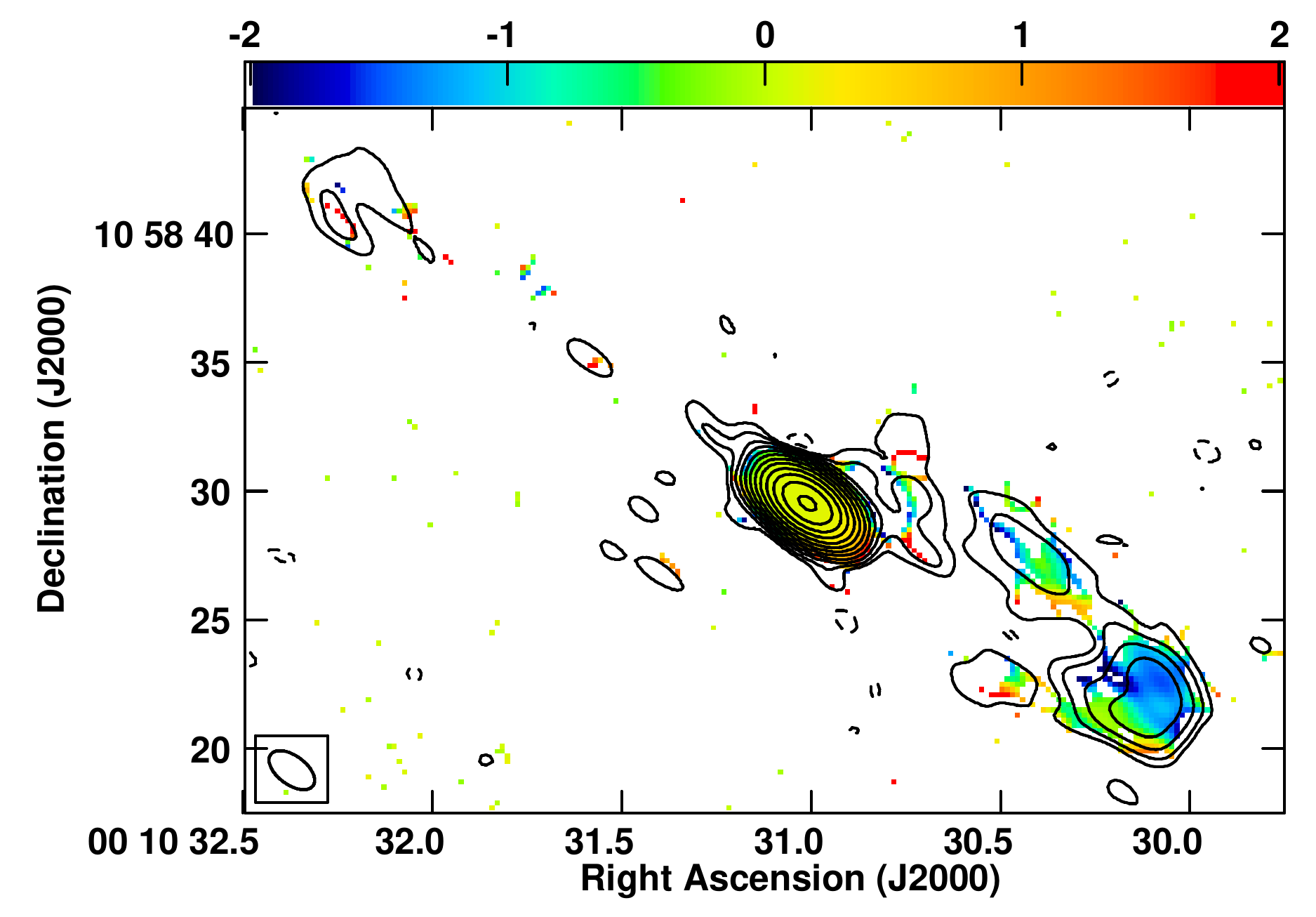}
\caption{\small VLA 6 GHz contour image of quasar PG0007+106 superimposed with red (top)
polarized intensity vectors and (bottom) in-band spectral index image. {  The inferred B-fields are  perpendicular to the polarization vectors assuming optically thin emission.}
The beam is $2.04\arcsec \times 1.12\arcsec$ with a PA of $53.54^\circ$. The peak surface brightness, $I_P$ is 123.98~mJy~beam$^{-1}$. The contour levels in percentage of the peak surface brightness $I_P$ are $(-0.045,~0.045,~0.09,~0.18,~0.35,~0.7,~1.4,~2.8,~5.6,~11.25,~22.5,~45,~90)$~Jy~beam$^{-1}$. The length of the EVPA vectors is proportional to polarized intensity with $5\arcsec$ corresponding to 0.156~mJy~beam$^{-1}$ \citep[see][]{Silpa2021}}.
 \label{fig2}
 \vspace{-27.44375pt}
\end{figure*}

This SSRQ ({\it aka} III~Zw~2) has been classified as a radio intermediate quasar by \citep{Miller1993,Falcke1996a,Falcke1996b}. It's host galaxy has prominent tidal tails and arms indicative of a recent merger systems \citep{Surace2001,Dunlop1993}. The source is known to be variable at both radio and optical wavelengths 
\citep[e.g.][]{Schnopper1978,Terasranta1992}. Using the 685~MHz uGMRT and 6 GHz VLA polarimetric observations, \citet{Silpa2021} were able to identify the signatures of a jet+wind-like stratification in the radio outflow of this PG quasar (see Figure~\ref{fig2}). The possible stratification in the outflow was revealed by the presence of different inferred B-field structures in the different frequency and different resolution images; uGMRT images sampled the outer sheath of the jet/lobe while the VLA images sampled the inner spine of the jet/lobe. The VLBI image from \citet{Wang2022} shows an unresolved core.

\subsubsection{PG1004+130} 
\begin{figure*}
\centering
\includegraphics[width=14.75cm,trim=0 0 0 0]{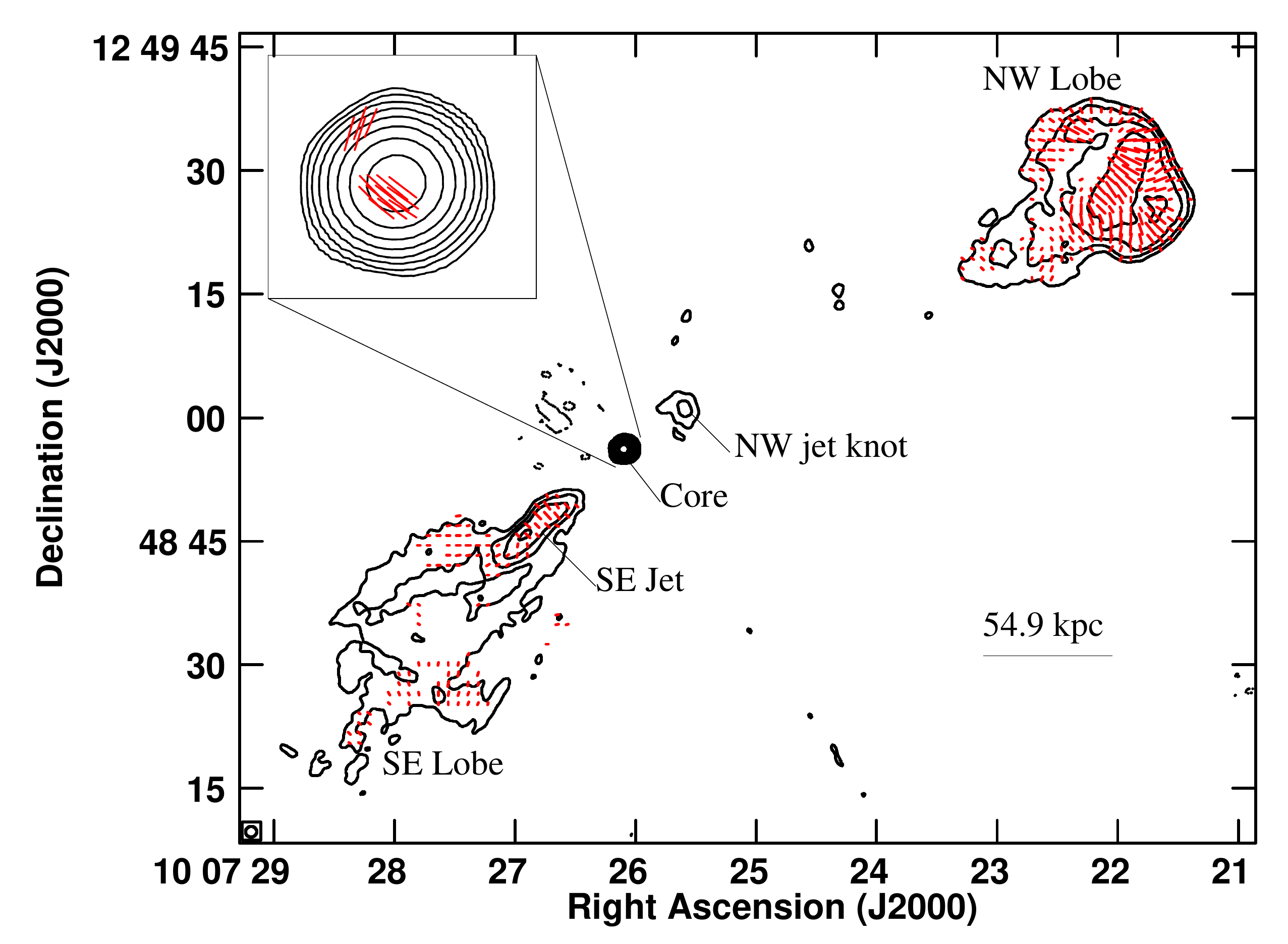}
\includegraphics[width=14.75cm]{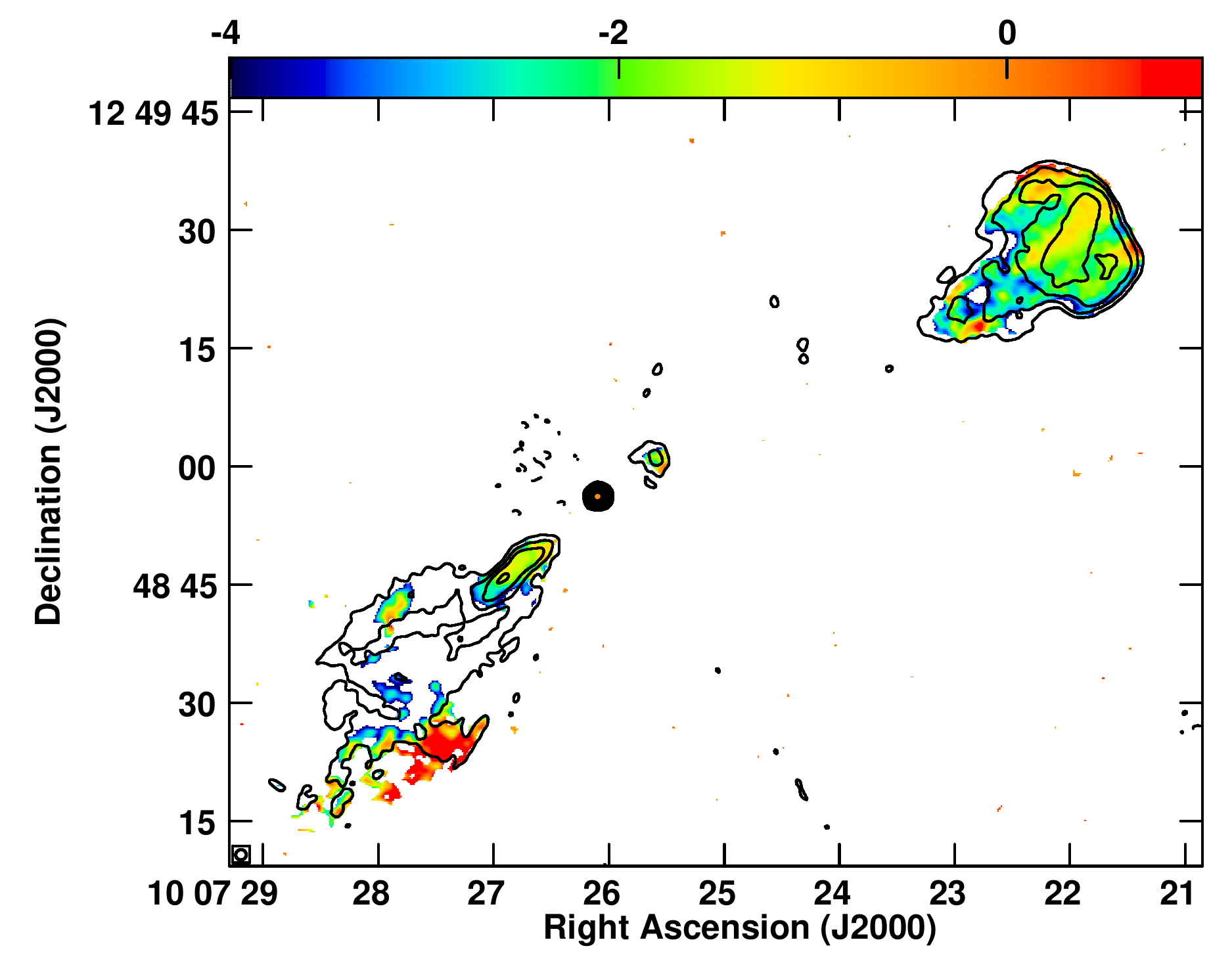}
\caption{\small VLA 6 GHz contour image of quasar PG1004+130 (4C 13.41) superimposed with red (top) 
polarized intensity vectors
and (bottom) in-band spectral index image.  
The beam is $1.35\arcsec \times 1.29\arcsec$ with a PA of $-80.03^\circ$. The peak surface brightness, $I_P$ is 28.62~mJy~beam$^{-1}$. The contour levels in percentage of the peak surface brightness $I_P$ are $(-0.4,~0.4,~1.0,~2.0,~4.0,~8.0,~16,~32,~64)$~Jy~beam$^{-1}$. The length of the EVPA vectors is proportional to polarized intensity with $10\arcsec$ corresponding to 2.27~mJy~beam$^{-1}$. The inset shows a zoomed-in image of the core and $0.5\arcsec$ corresponding to 0.05~mJy~beam$^{-1}$. 
}
\label{fig3}
\vspace{-8.48894pt}
\end{figure*}

This SSRQ ({\it aka} 4C+13.41) appears to be a restarted source as observed in the bright jet-core-counterjet emission on the 30$\arcsec$-scales (Figure~\ref{fig3}). A clear surface brightness discontinuity is observed between the jet-counterjet and the larger lobe emission. The inferred B-field is transverse to the jet direction for the optically thick radio core. The terminal region on the western side does not show a typical B-field configuration in its hotspot (i.e., transverse to the jet direction). This could indicate weakening of the hotspot as the previous episode of activity has died down. The jet seen inside the south eastern lobe shows B-field vectors parallel to the jet direction (assuming optically thin emission) ending in a weak hotspot with B-field vectors nearly perpendicular to the jet direction. This B-field structure is similar to the one seen in 3C219 by \citet{Clarke1992} for its restarting jet. The south eastern lobe shows diffuse emission that is not well detected in our higher resolution images. { The VLBA image from \citet{Wang2022} indicates that the parsec-scale (approaching) jet is aligned well with the south-eastern inner jet.}
 
\citet{Gopal-Krishna2000} have classified this quasar as a hybrid FRI/FRII morphology source { and \citet{Miller2006} found an X-ray counterpart a little upstream of the radio FR I jet, with a flat photon index and a concave SED more typical of FR II quasar jets than FR I jets. \citet{Gopal-Krishna2000} have explained this hybrid nature as resulting from environmental differences in density, i.e. the density of the surrounding medium is higher towards the SE causing decollimation.} This source is also a broad absorption line (BAL) QSO \citep{Wills1999}.

\subsubsection{PG1048$-$090} 
\begin{figure*}
\centering
\includegraphics[width=14.25cm,trim=0 0 0 0]{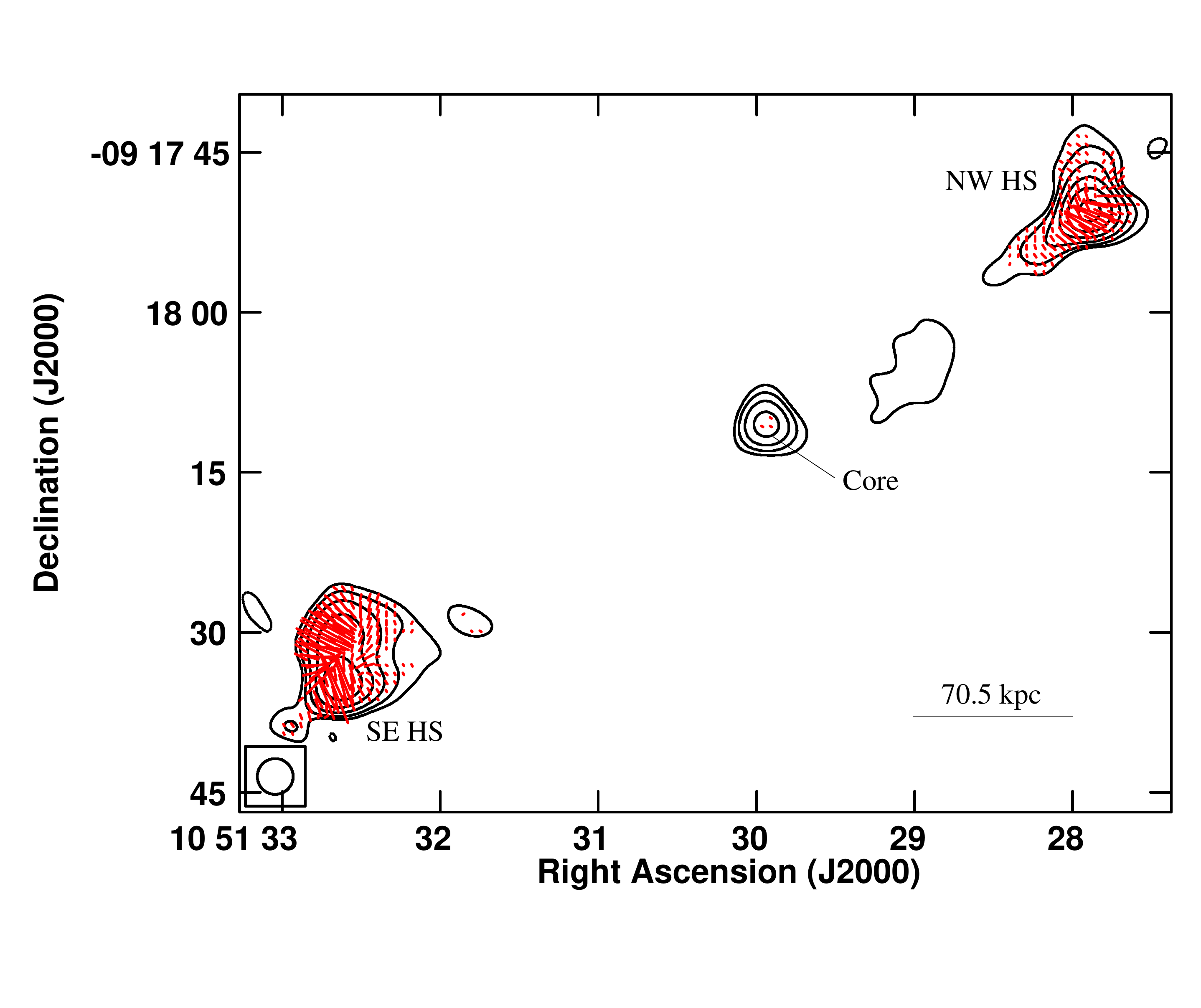}
\includegraphics[width=14.25cm]{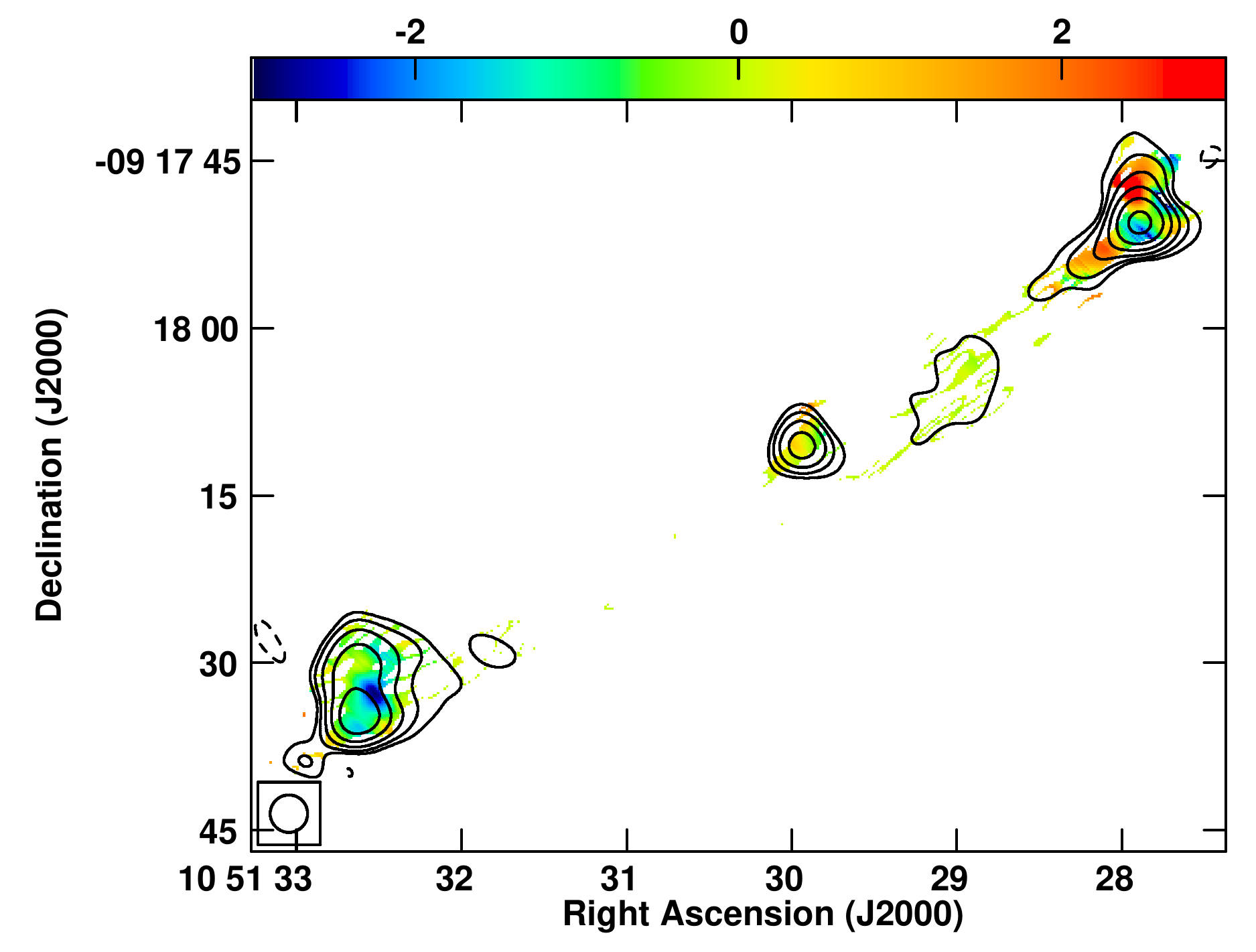}
\caption{\small VLA 6 GHz contour image of quasar PG1048-090 (3C246) superimposed with red (top) polarized intensity vectors and (bottom) in-band spectral index image. 
 The beam is $3.35\arcsec \times 3.35\arcsec$ with a PA of $0^\circ$. The peak surface brightness, $I_P$ is 0.1836~Jy~beam$^{-1}$. The contour levels in percentage of the peak surface brightness $I_P$ are $(-2.5,~2.5,~5.0,~10.0,~20.0,~40.0,~80.0)$~Jy~beam$^{-1}$. The length of the EVPA vectors is proportional to polarized intensity with $5\arcsec$ corresponding to 6.25~mJy~beam$^{-1}$. 
 }
 \label{fig4}
 \vspace{-54.23109pt}
 \end{figure*}

This SSRQ ({\it aka} 3C246) is a member of a galaxy cluster \citep{Yates1989}. It is relatively poorly imaged at radio frequencies compared to the other sources. Previous VLA images show the presence of a core and two hotspots; however the jet and lobe emission is very faint \citep{Miller1993, Dennett2000, Kellermann1994}. We find a similar structure in our images (Figure~\ref{fig4}). {  The Y-shaped ears around the north-western hotspot are due to deconvolution errors that remain in spite of self-calibration and must be regarded as image artefacts.} We find that the core and two hotspots show extensive polarization; however their fractional polarization is low. This would be consistent with the presence of X-ray emitting gas in the galaxy cluster \citep[e.g.,][]{Sarazin1986}. The inferred B-field is transverse to the jet direction for the optically thick radio core. The western hotspot shows a polarization structure (B-field aligned along the jet direction) consistent with the jet entering the hotspot region. The in-band spectral index image (Figure~\ref{fig4}) shows that the core has an inverted spectrum while the hotspots have steep spectra. { The VLBA image from \citet{Wang2022} indicates that the parsec-scale (approaching) jet is in the north-west direction.}

\subsubsection{PG1100+772} 
\begin{figure*}
\centering
\includegraphics[width=18.00cm,trim=0 0 0 0]{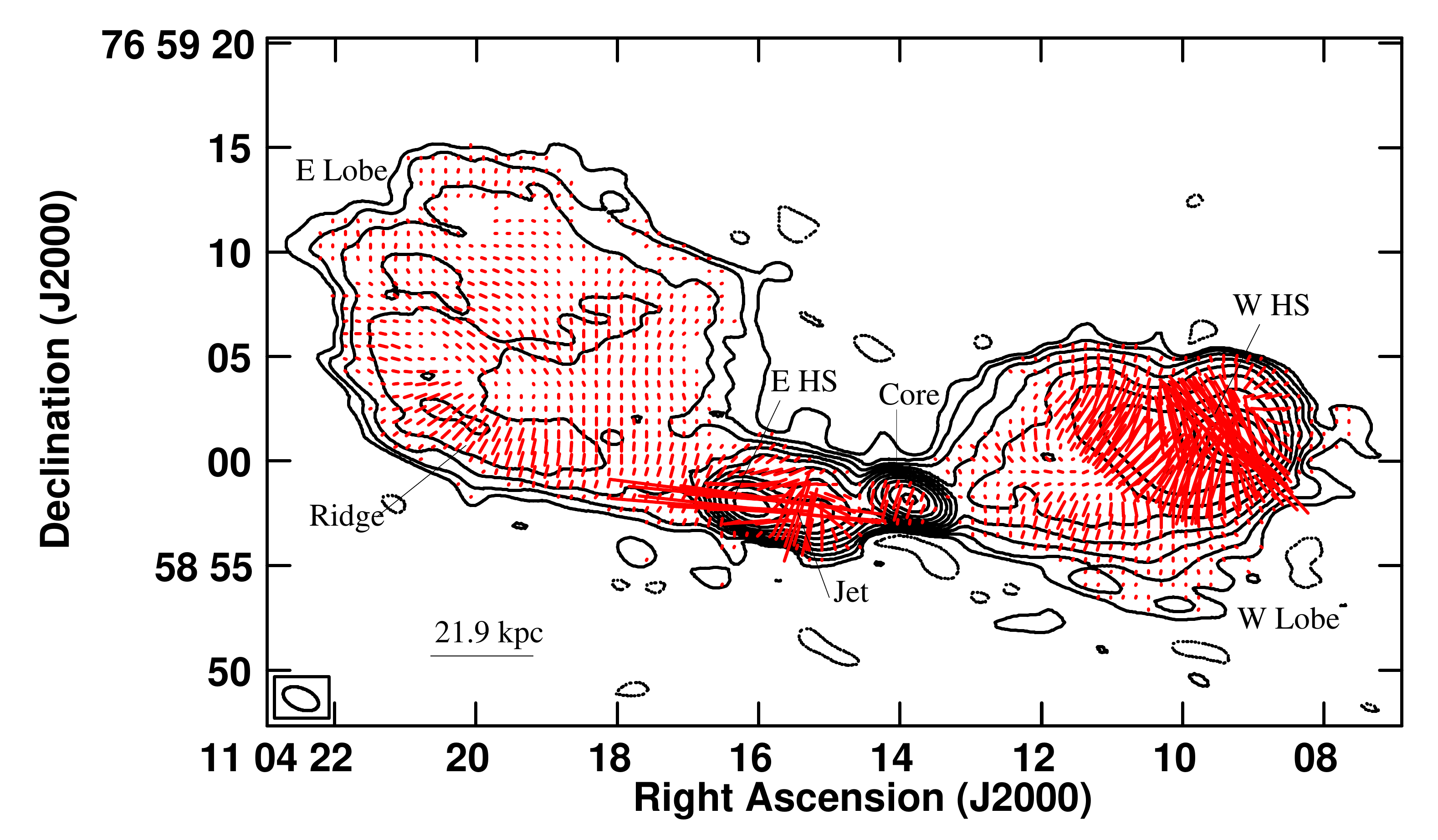}
\includegraphics[width=18.00cm]{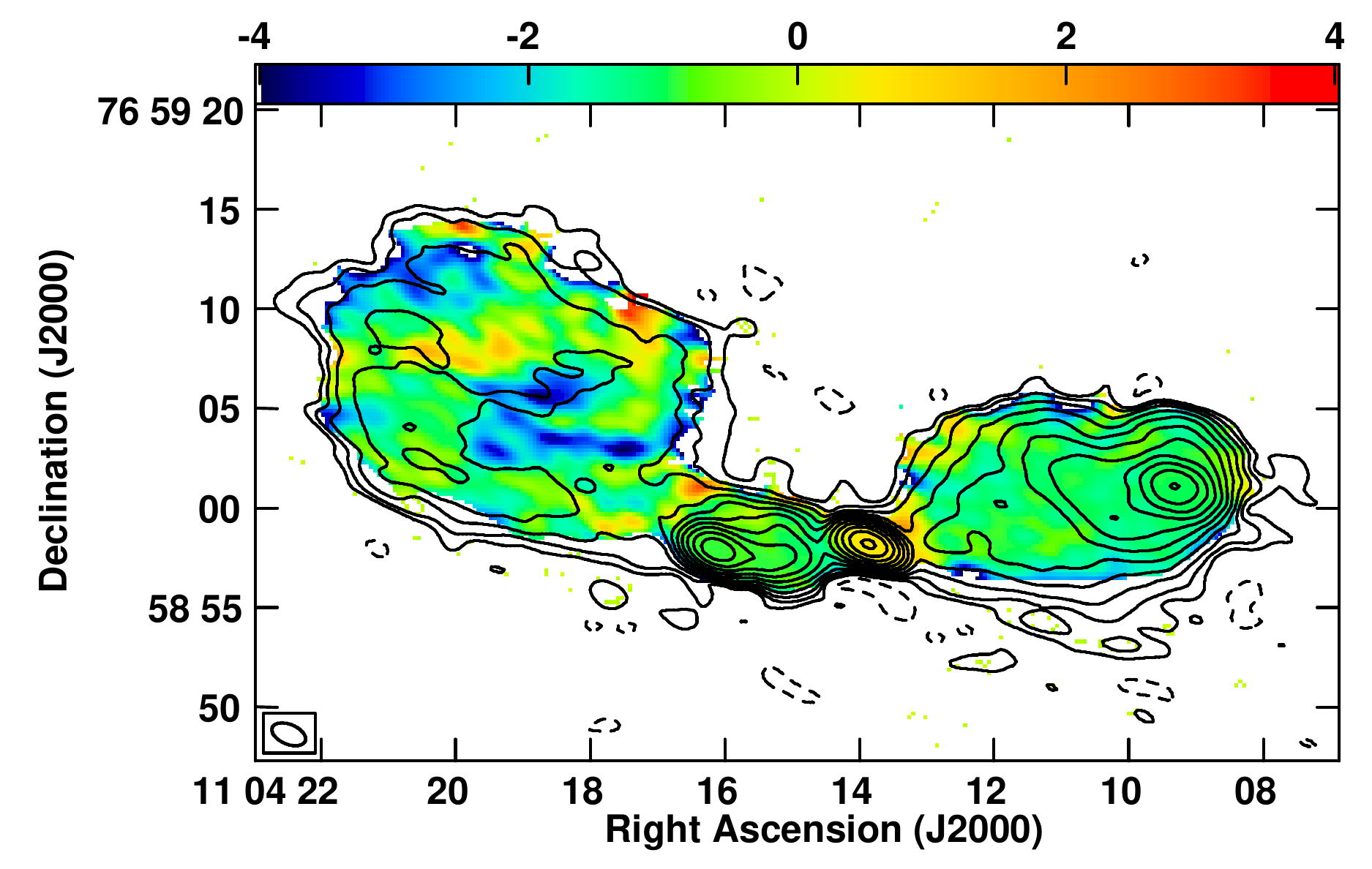}
\caption{\small VLA 6 GHz contour image of quasar PG1100+772 superimposed with red (top) 
polarized intensity vectors and (bottom) in-band spectral index image. {  The inferred B-fields are  perpendicular to the polarization vectors assuming optically thin emission.}
The beam is $1.79\arcsec \times 1.00\arcsec$ with a PA of $67.35^\circ$. The peak surface brightness, $I_P$ is 107.6~mJy~beam$^{-1}$. The contour levels in percentage of the peak surface brightness $I_P$ are $(-0.09,~0.09,~0.18,~0.35,~0.7,~1.4,~2.8,~5.6,~11.25,~22.5,~45,~90)$~Jy~beam$^{-1}$. The length of the EVPA vectors is proportional to polarized intensity with $5\arcsec$ corresponding to 3.57~mJy~beam$^{-1}$. 
}
\label{fig5}
\vspace{-24.94544pt}
\end{figure*}

\begin{figure*}
\centering
\includegraphics[width=14.55cm,trim=0 0 0 0]{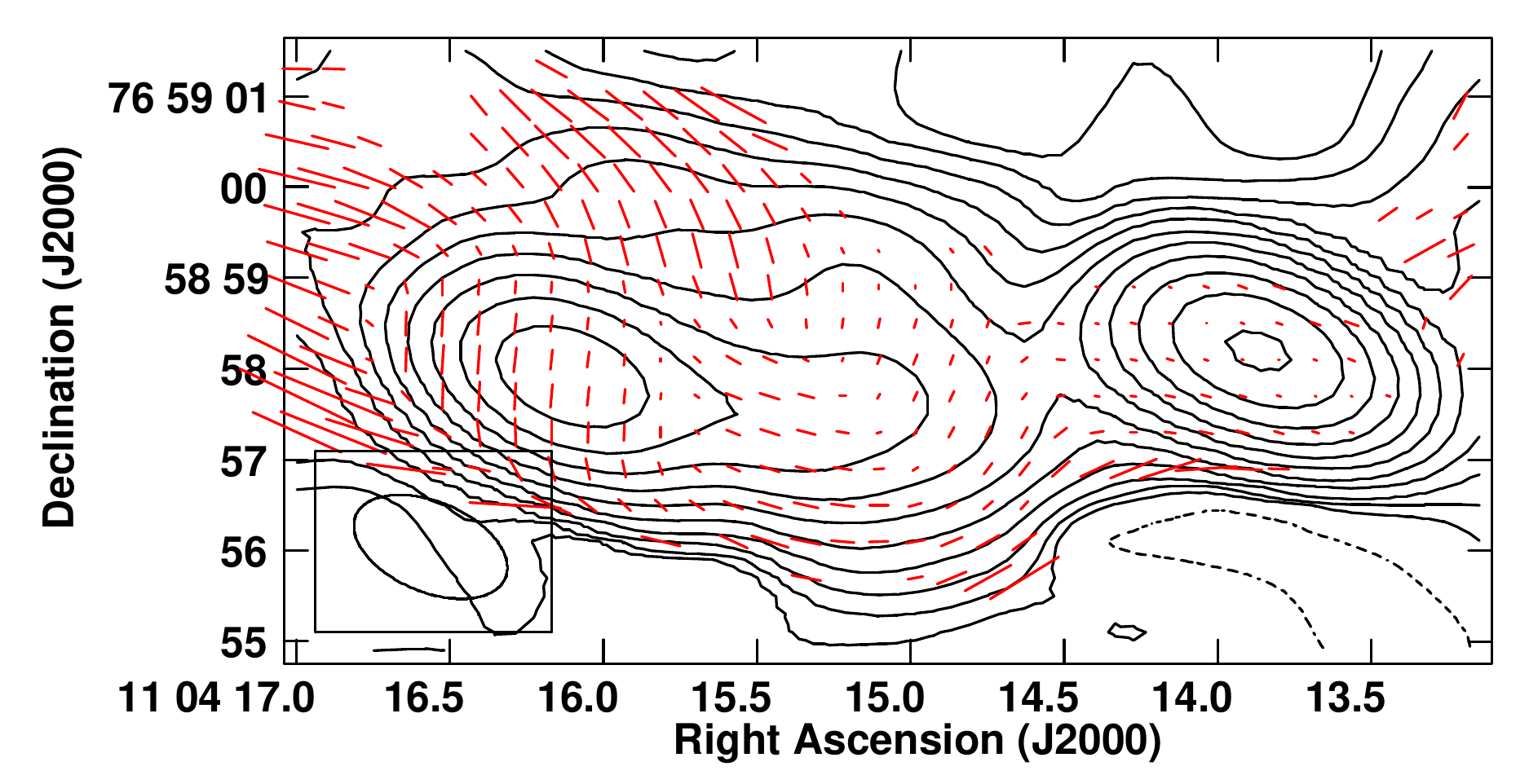}
\caption{\small {  VLA 6 GHz contour image of the core and eastern hotspot of quasar PG1100+772 (3C249.1) superimposed with red polarization vectors whose length is proportional to fractional polarization, rotated by 90$^\circ$ to indicate the B-field direction, assuming optically thin emission. The beam is $1.79\arcsec \times 1.00\arcsec$ with a PA of $67.35^\circ$. The peak surface brightness, $I_P$ is 107.6~mJy~beam$^{-1}$. The contour levels in percentage of the peak surface brightness $I_P$ are $(-0.09,~0.09,~0.18,~0.35,~0.7,~1.4,~2.8,~5.6,~11.25,~22.5,~45,~90)$~Jy~beam$^{-1}$. 1$\arcsec$ length of the vector corresponds to 25\%.}}
\label{fig5.1}
\end{figure*}

This SSRQ ({\it aka} 3C249.1) is a candidate restarted source \citep[][Figure~\ref{fig5}]{Marecki2012,Fernini2007}. The western lobe is FRII-like while the eastern one does not have a terminal hotspot and is FRI-like. This makes the source a hybrid FRI/FRII radio source. The inferred B-field is transverse to the jet direction for the optically thick radio core. The polarization features in the inner ``hotspot'' are consistent with a terminal shock in that the B-vectors are transverse to the jet direction here { (Figure~\ref{fig5.1})}. Further down the eastern lobe, the B-field vectors resemble the bubble-like lobe structures observed in several Seyfert galaxies \citep{Kharb2006,Sebastian2020} and is relatively flat in spectral index. { The VLBI image from \citet{Wang2022} shows an unresolved core.}

This source was observed at VLA 5~GHz B/C array by \citet{Bridle1994} { and \citet{Gilbert2004}} who identified that the two lobes are asymmetric in size, intensity and surface brightness. The west lobe is edge-brightened with a well resolved hotspot near its western end. The east lobe is edge-darkened except for a resolved ridge along its southern boundary. \citet{Bridle1994} note that a jet enters it on its extreme southern edge and terminates abruptly at $7.7\arcsec$ ($\sim$34~kpc) from the central feature. The degree of polarization is greatest at the edges of the lobes, particularly east of eastern hotspot at the end of the jet, where it reaches 55\% to 65\%. Interestingly, we have detected a polarization of 44$\pm$4\% in this region which is a $\sim$10 - 40\% reduction. The EVPA vectors are roughly perpendicular to the main ridge in the east lobe and to the outer boundaries of the west lobe. At 0.35$\arcsec$ resolution from the combined A, B, and C configuration data, all structures in the east lobe are fully resolved. The jet bends at least twice on its path to the resolved terminal hot spot at its east end. 

{ Many different explanations have been proposed in the literature for the  peculiar morphology of PG1100+772, including a dense cluster environment \citep{Hintzen1984}, cooling flows \citep{Crawford1988}, interacting systems \citep{Stockton1983}, the “flip-flop” mechanism \citep{Lonsdale1983} and restarted activity \citep{Marecki2012}.} There is an emission line system extending for several arcseconds around the quasar, plus an extended optical continuum that may be of stellar origin \citep{Richstone1977,Boroson1984}.

\begin{figure*}
\centering
\includegraphics[width=14.55cm,trim=0 0 0 0]{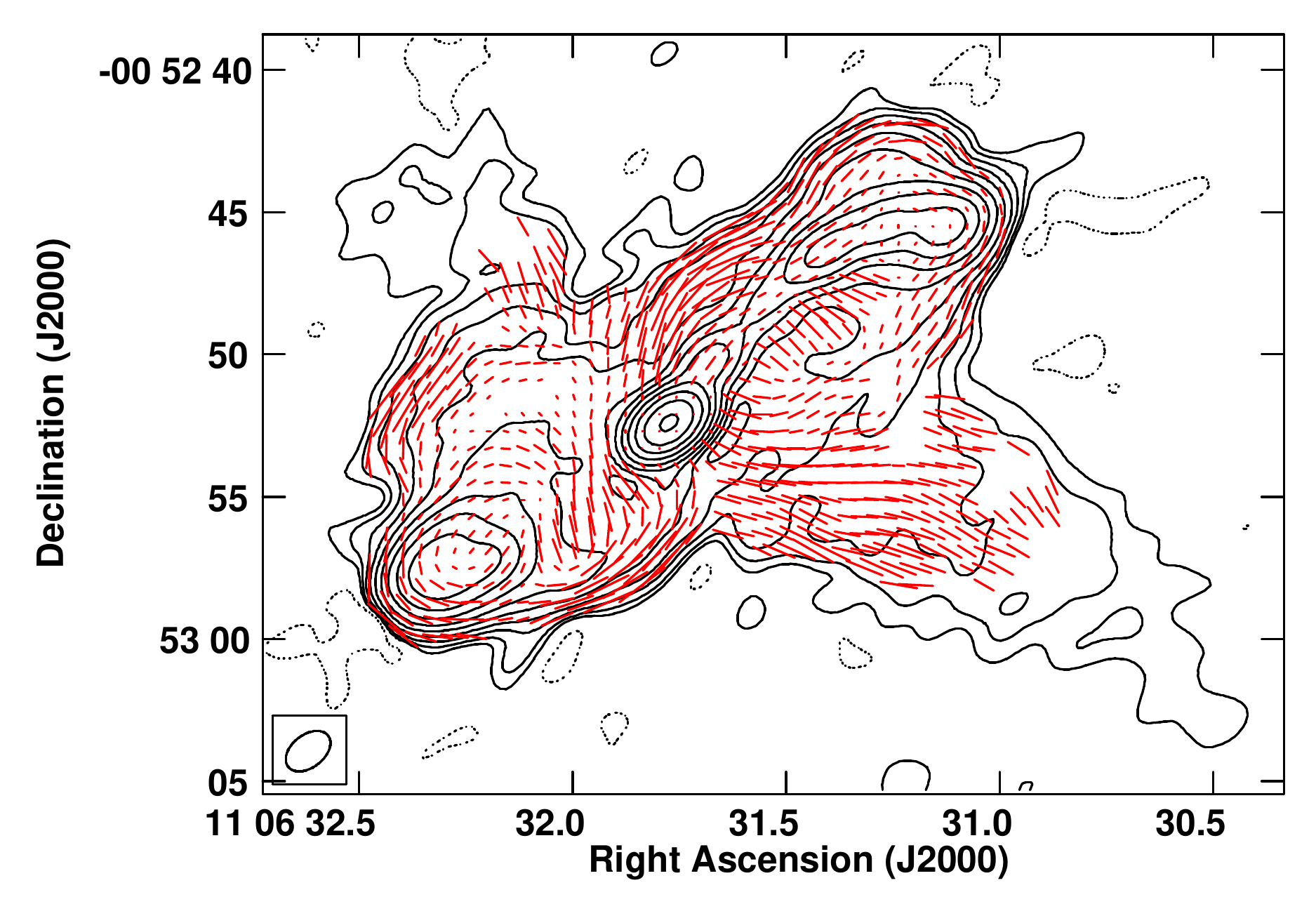}
\caption{\small {  VLA 6 GHz contour image of PG1103$-$006 (4C-00.43) superimposed with red polarization vectors whose length is proportional to fractional polarization, rotated by 90$^\circ$ to indicate the B-field direction, assuming optically thin emission. The beam is $1.79\arcsec \times 1.15\arcsec$ with a PA of $-51.79^\circ$. The peak surface brightness, $I_P$ is 128.7~mJy~beam$^{-1}$. The contour levels in percentage of the peak surface brightness $I_P$ are $(-0.045,~0.045,~0.09,~0.18,~0.35,~0.7,~1.4,~2.8,~5.6,~11.25,~22.5,~45,~90)$~Jy~beam$^{-1}$. 2$\arcsec$ length of the vector corresponds to 62.5\%.}}
\label{fig6.1}
\end{figure*}

\begin{figure*}
\centering
\includegraphics[width=16cm,trim=0 0 0 0]{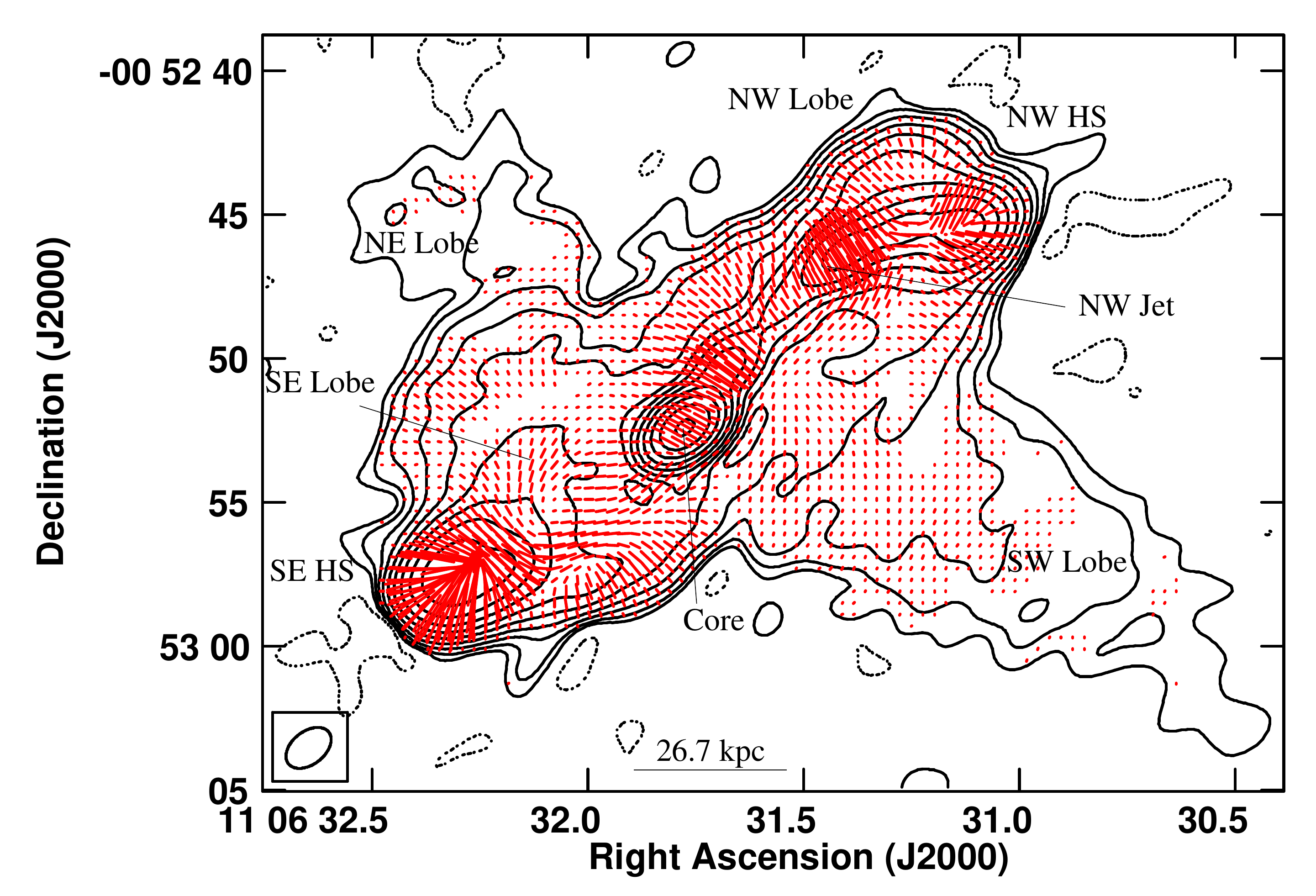}
\includegraphics[width=16cm]{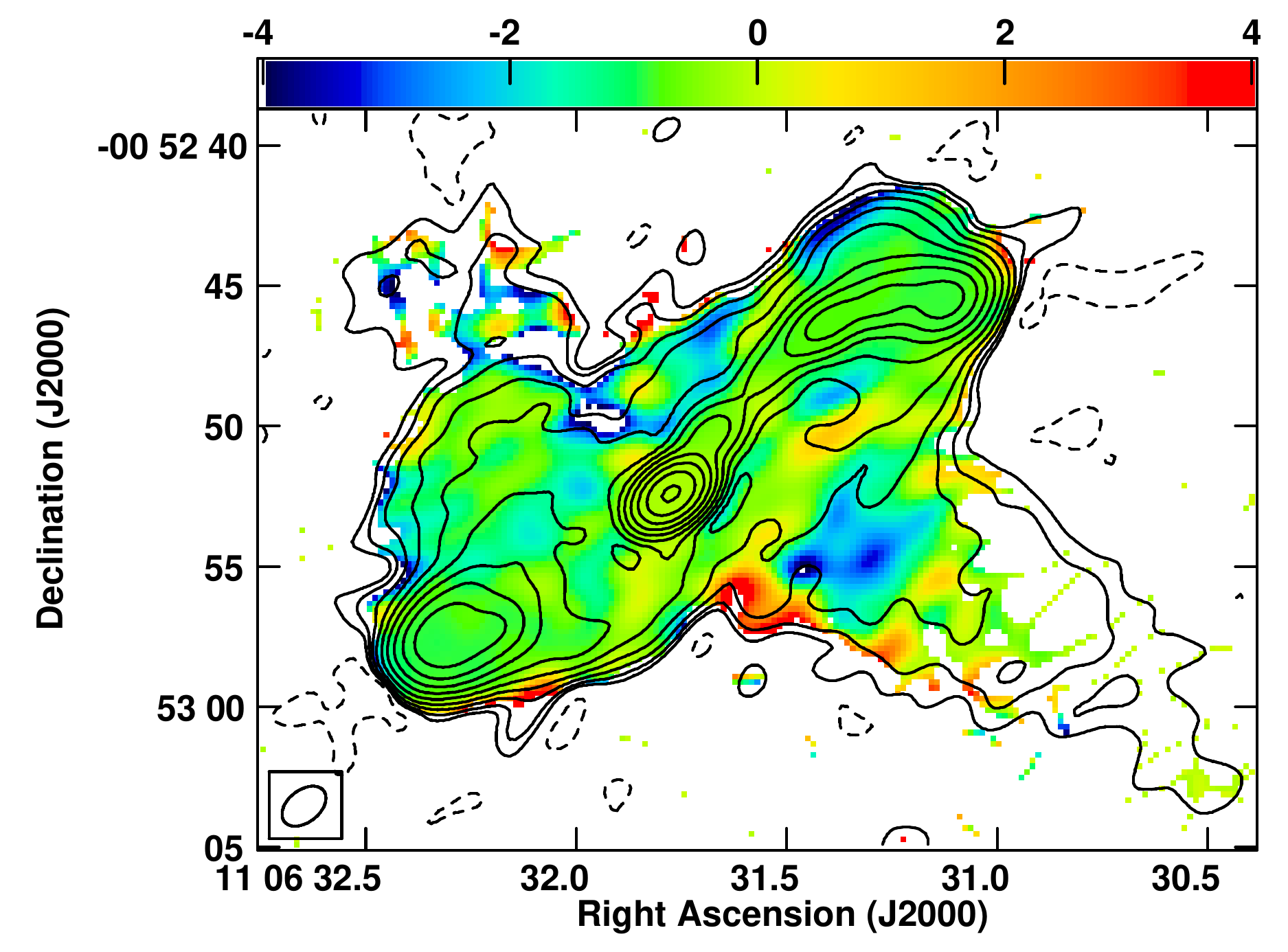}
\caption{\small VLA 6 GHz contour image of quasar PG1103$-$006 superimposed with red (top) 
polarized intensity vectors
and (bottom) in-band spectral index image.
The beam is $1.79\arcsec \times 1.15\arcsec$ with a PA of $-51.79^\circ$. The peak surface brightness, $I_P$ is 128.7~mJy~beam$^{-1}$. The contour levels in percentage of the peak surface brightness $I_P$ are $(-0.09,~0.09,~0.18,~0.35,~0.7,~1.4,~2.8,~5.6,~11.25,~22.5,~45,~90)$~Jy~beam$^{-1}$. The length of the EVPA vectors is proportional to polarized intensity with $2\arcsec$ corresponding to 2.00~mJy~beam$^{-1}$. 
}
\label{fig6}
\vspace{-19.01599pt}
\end{figure*}

\subsubsection{PG1103$-$006} This RL quasar ({\it aka} { 4C -00.43}) is an X-shaped radio source (Figure~\ref{fig6.1},\ref{fig6}). The jets appear to be S-shaped, especially going North, which could be consistent with jet precession. This quasar has been observed previously by the VLA by \citet{Hutchings1998} who did not detect the X-shape as prominently. \citet{Hutchings1996} studied this source as those residing inside galaxy cluster environments. \citet{Rector1995} did not see any correlation between galaxy density and quasar radio morphology in their sample of QSOs of which PG1103$-$006 is one. Such a correlation is expected from jet-ICM interaction. Our polarization structures in the inner lobes match those of \citet{Reid1999}; however, they also did not pick up the X-shaped lobe emission in their VLA 4.9 GHz B-array images. { The VLBI image from \citet{Wang2022} shows that the parsec-scale (approaching) jet is aligned with the north-west jet direction.}

The inferred B-field structure in the north-western jet stays aligned with the jet throughout, all the way to the north-western hotspot. The inferred B-field in the radio core is transverse to the jet direction for the optically thick core. The inferred B-field appears to be be bow-shock-like in both the hotspots, as seen in the southern hotspot of III~Zw~2. Extensive polarization is observed in the X-shaped wings. { Although the inferred B-fields are aligned with the direction of the wings { (see Figure~\ref{fig6.1})}, the vectors do not follow the flow ``around the corner'' unambiguously. Nevertheless, we do find them to be largely} consistent with the ``hydrodynamical backflow model'' \citep[see][]{Cotton2020}. We discuss the models for X-shaped sources in section~\ref{Disc} ahead. 

\subsubsection{PG1226+023} 
\begin{figure*}
\centering
\includegraphics[width=11.5cm,trim=0 0 0 0]{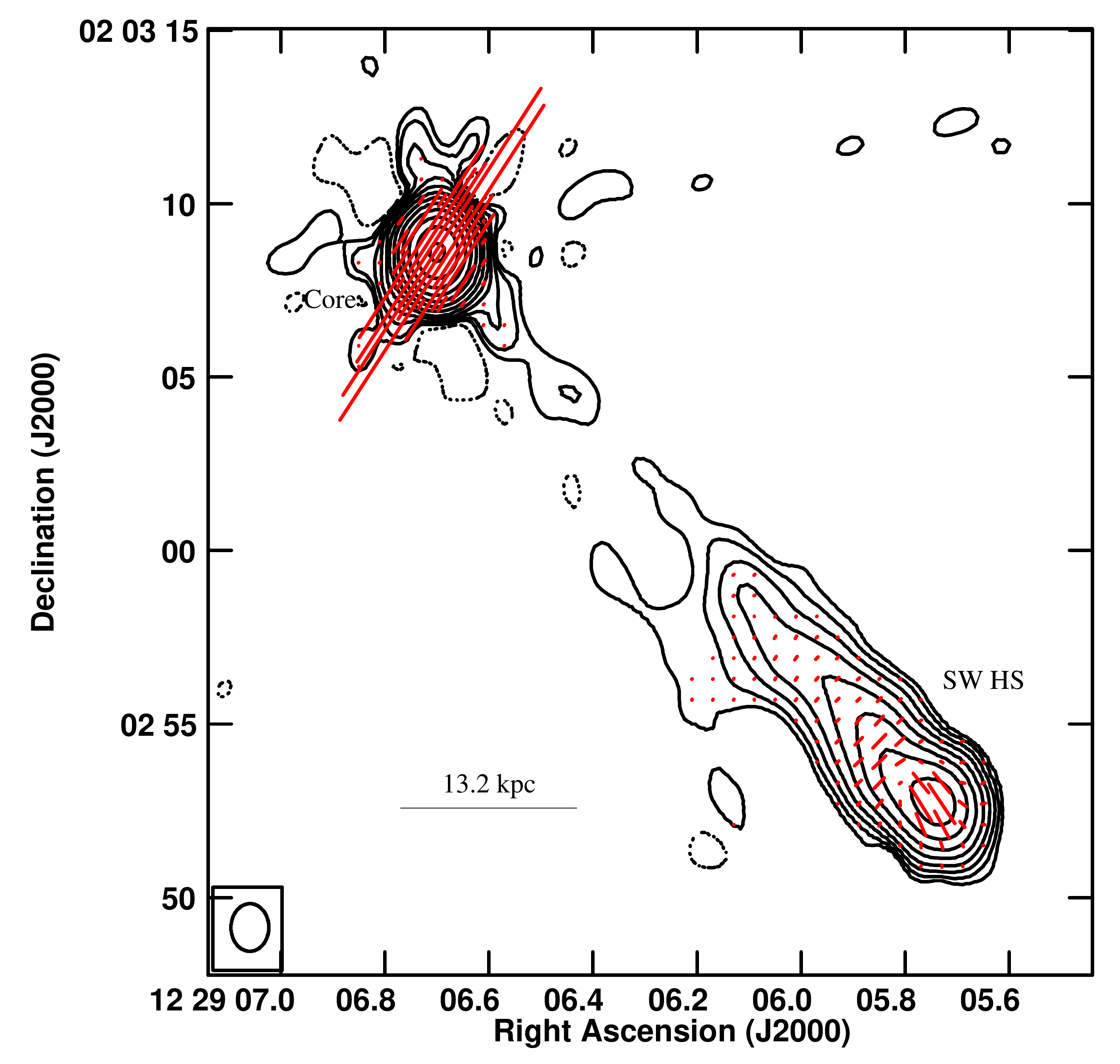}
\includegraphics[width=11.5cm]{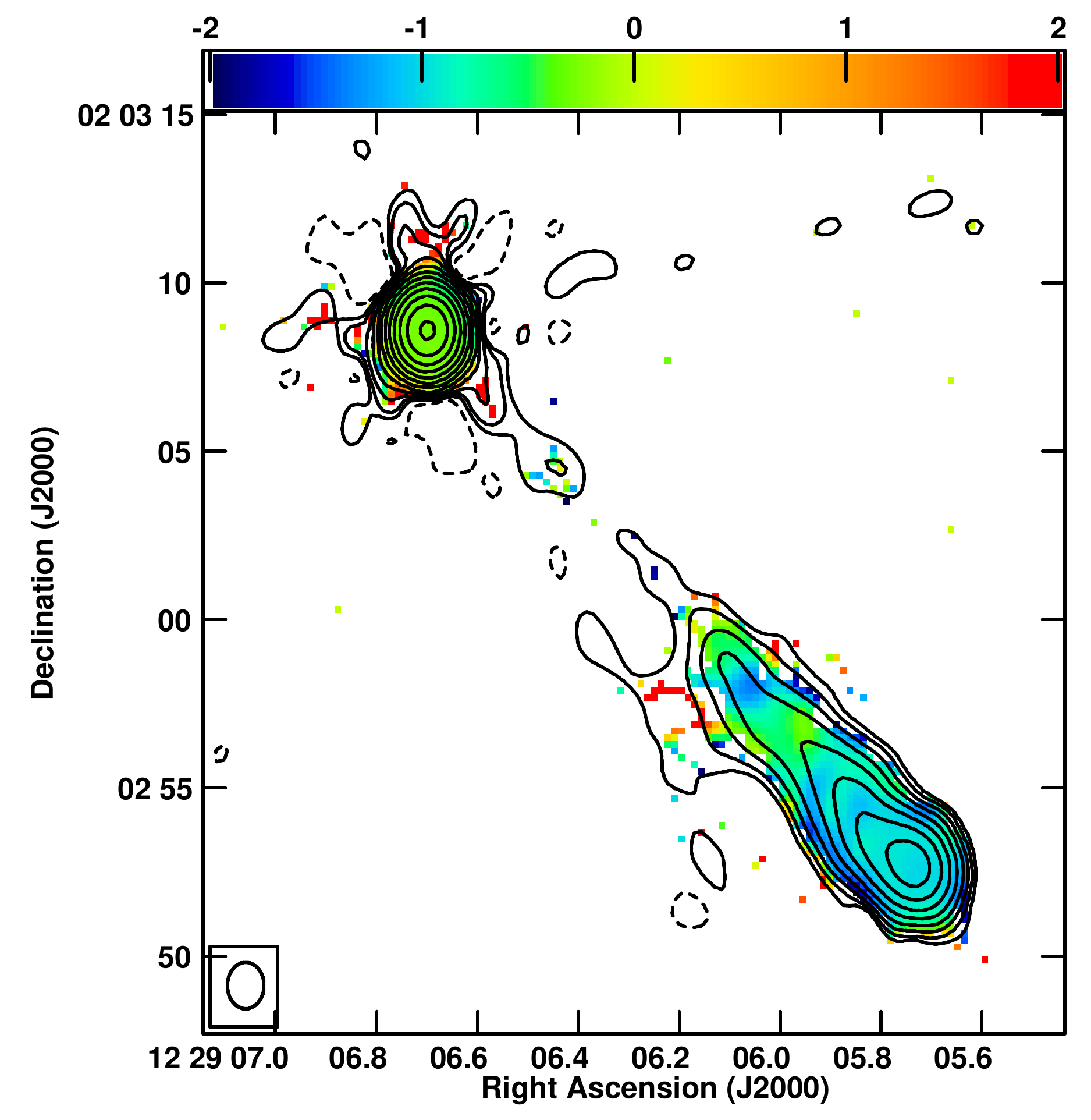}
\caption{\small VLA 6 GHz contour image of quasar PG1226+023 (3C273) superimposed with red (top) polarized intensity vectors 
and (bottom) in-band spectral index image.
The beam is $1.37\arcsec \times 1.09\arcsec$ with a PA of $-0.43^\circ$. The peak surface brightness, $I_P$ is 21.05~Jy~beam$^{-1}$. The contour levels in percentage of the peak surface brightness $I_P$ are $(-0.045,~0.045,~0.09,~0.18,~0.35,~0.7,~1.4,~2.8,~5.6,~11.25,~22.5,~45,~90)$~Jy~beam$^{-1}$. The length of the EVPA vectors is proportional to polarized intensity with $2\arcsec$ corresponding to 333.33~mJy~beam$^{-1}$.
}
\label{fig7}
\vspace{-19.45554pt}
\end{figure*}

This famous FSRQ ({\it aka} 3C273) is the first identified quasar \citep{Schmidt1963}. It is one of the closest and most luminous of all quasars. Imaging of this source has been done at multiple wavebands \citep[e.g.,][]{Perley2017}. 3C273 has a bright, flat-spectrum nucleus with highly variable flux density, and a one-sided, highly polarized, narrow jet extending $\sim23\arcsec$ to the southwest of the nucleus. Our 5~GHz image (Figure~\ref{fig7}) is consistent with the previous total intensity, polarization and spectral index images of 3C273 \citep{Perley2017}. The inferred B-field is transverse to the jet direction in the optically thick radio core, i.e., the unresolved base of the jet. The inferred B-field is aligned with the jet direction along the jet and becomes transverse to it at the position of the southern hotspot. { The VLBI image from \citet{Wang2022} indicates that the parsec-scale (approaching) jet is towards the south-west and shows clear curvature.}

\subsubsection{PG1309+355}
\begin{figure*}
\centering
\includegraphics[width=13cm,trim=0 0 0 0]{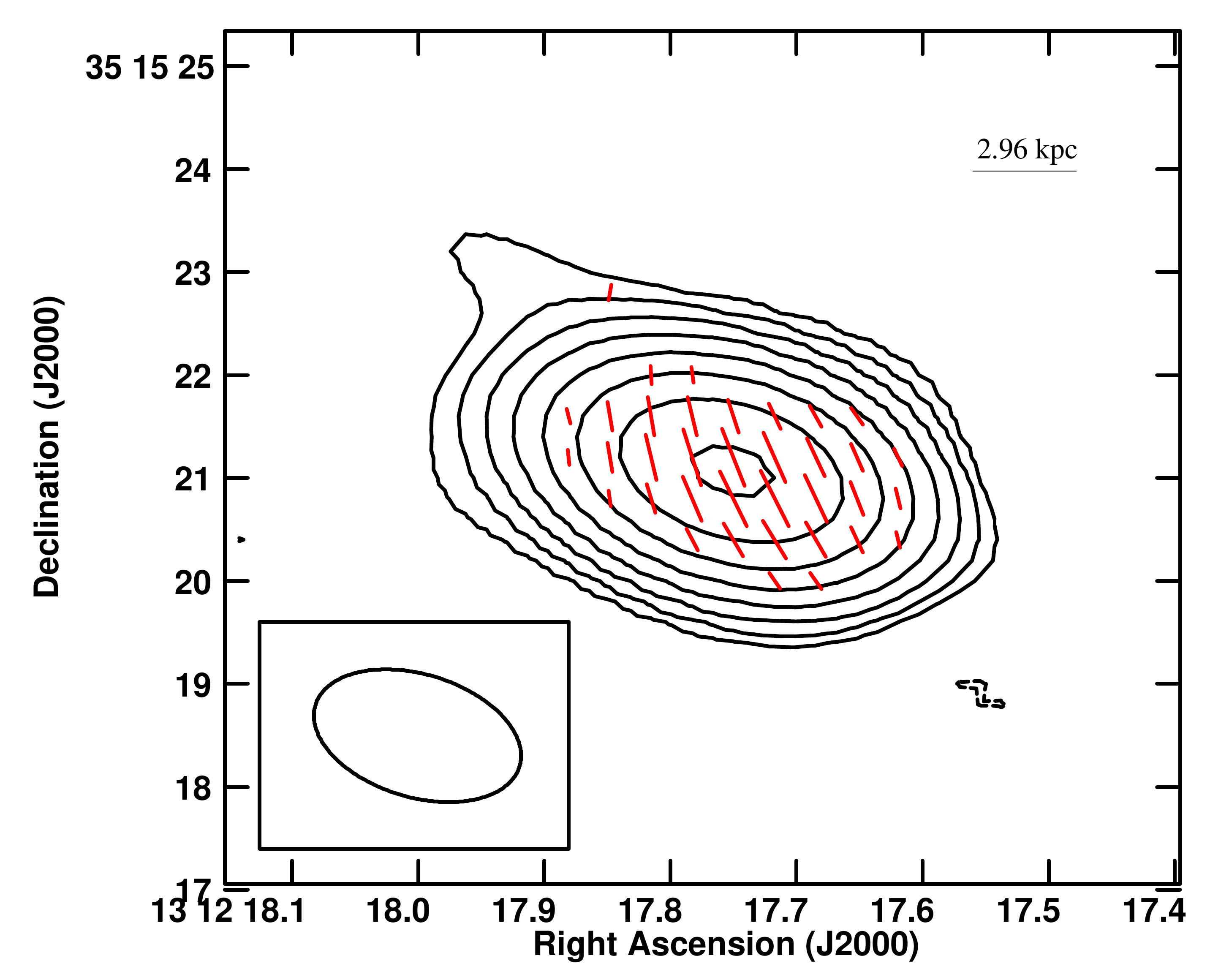}
\includegraphics[width=13cm]{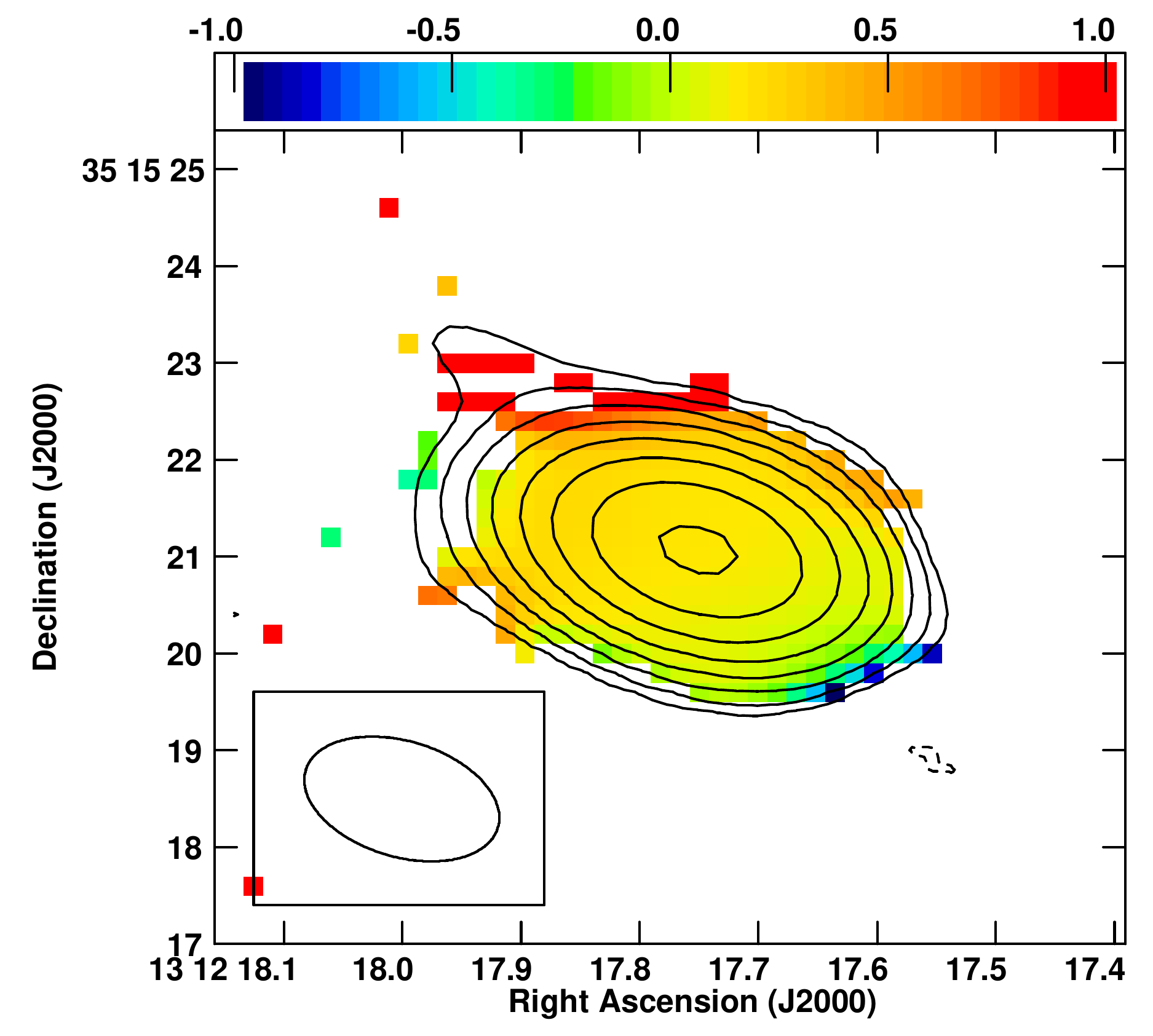}
\caption{\small VLA 6 GHz contour image of quasar PG1309+355 (Ton 1565) superimposed with red (top) polarized intensity vectors and (bottom) in-band spectral index image.
The beam is $2.07\arcsec \times 1.19\arcsec$ with a PA of $73.31^\circ$. The peak surface brightness, $I_P$ is 0.047415~Jy~beam$^{-1}$. The contour levels in percentage of the peak surface brightness $I_P$ are $(-0.7,~0.7,~1.4,~2.8,~5.6,~11.25,~22.5,~45,~90)$~Jy~beam$^{-1}$. The length of the EVPA vectors is proportional to polarized intensity with $1.0\arcsec$ corresponding to 0.23~mJy~beam$^{-1}$. 
 }
 \label{fig8}
 \vspace{-4.4572pt} 
 \end{figure*}

This radio-intermediate FSRQ ({\it aka} Ton1565) has been suggested to be a boosted RQ quasar \citep{Miller1993}. It shows a compact radio morphology in VLA A and D array 5~GHz data \citep{Kellermann1994}. Its host galaxy shows evidence of a spiral-like substructure \citep{Kim2008}. In view of its small viewing angle, strong UV absorption lines are observed in its HST spectrum \citep{Brandt2000}. \citet{Brandt2000} note that it is moderately X-ray weak for a radio-loud QSO. We had chosen this object as being potentially extended based on its NVSS image. However, we detect only a compact radio core in our 5~GHz data (Figure~\ref{fig8}) which shows marginal fractional polarization ($\sim0.4$\%). Our VLA data has failed to detect any extended diffuse emission or hotspots on kpc-scales. An archival VLA A-array image\footnote{http://www.vla.nrao.edu/astro/archive/pipeline/position/J131217.7+351521/} at 22~GHz (resolution $100\times75$~mas) does indicate a faint jet-like extension to the west of the core. If this is indeed the kpc-scale jet direction in PG1309+355, the inferred B-field is transverse to the jet direction for optically thick core emission. { The VLBI image from \citet{Wang2022} shows an unresolved core.}

The in-band spectral index (Figure~\ref{fig8}) also suggests a gradient to the south-west which could hint at the presence of an unresolved core-jet structure. Its EVN image, however, suggests a mas-scale core-jet structure to the south-east \citep{Falcke1996a}. It is possible that the jet curves towards the south-west on scales intermediate between the VLA and EVN+MERLIN. Large jet bents going from parsec- to kpc-scale are commonly observed in blazars \citep{Kharb10}. Moreover, \citet{Falcke1996b} suggest that PG1309+355 might not be a boosted radio-quiet quasar based on their observations.

\subsubsection{PG1704+608} 
\begin{figure*}
\centering
\includegraphics[width=9.20cm,trim=0 0 0 0]{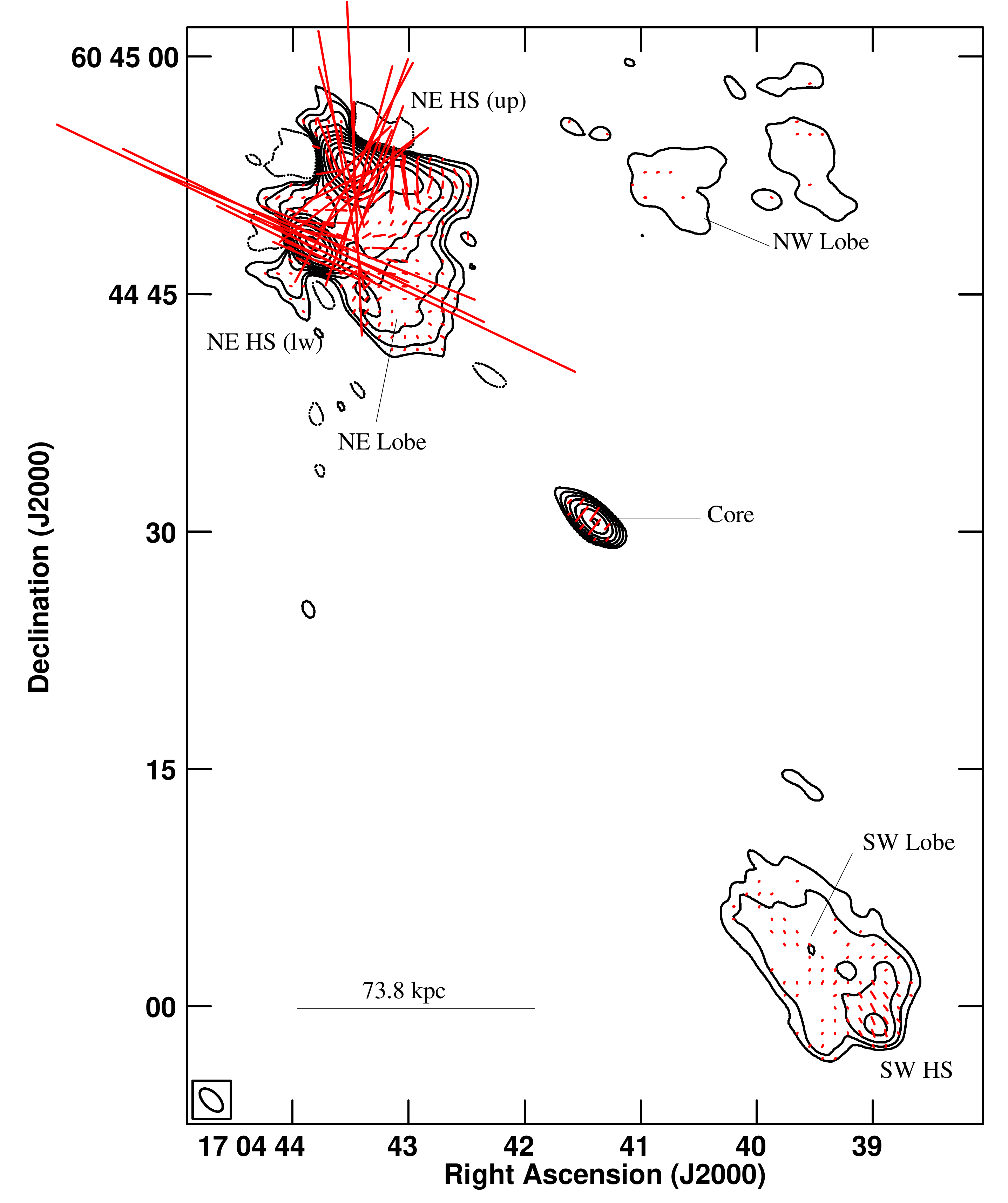}
\includegraphics[width=9.20cm]{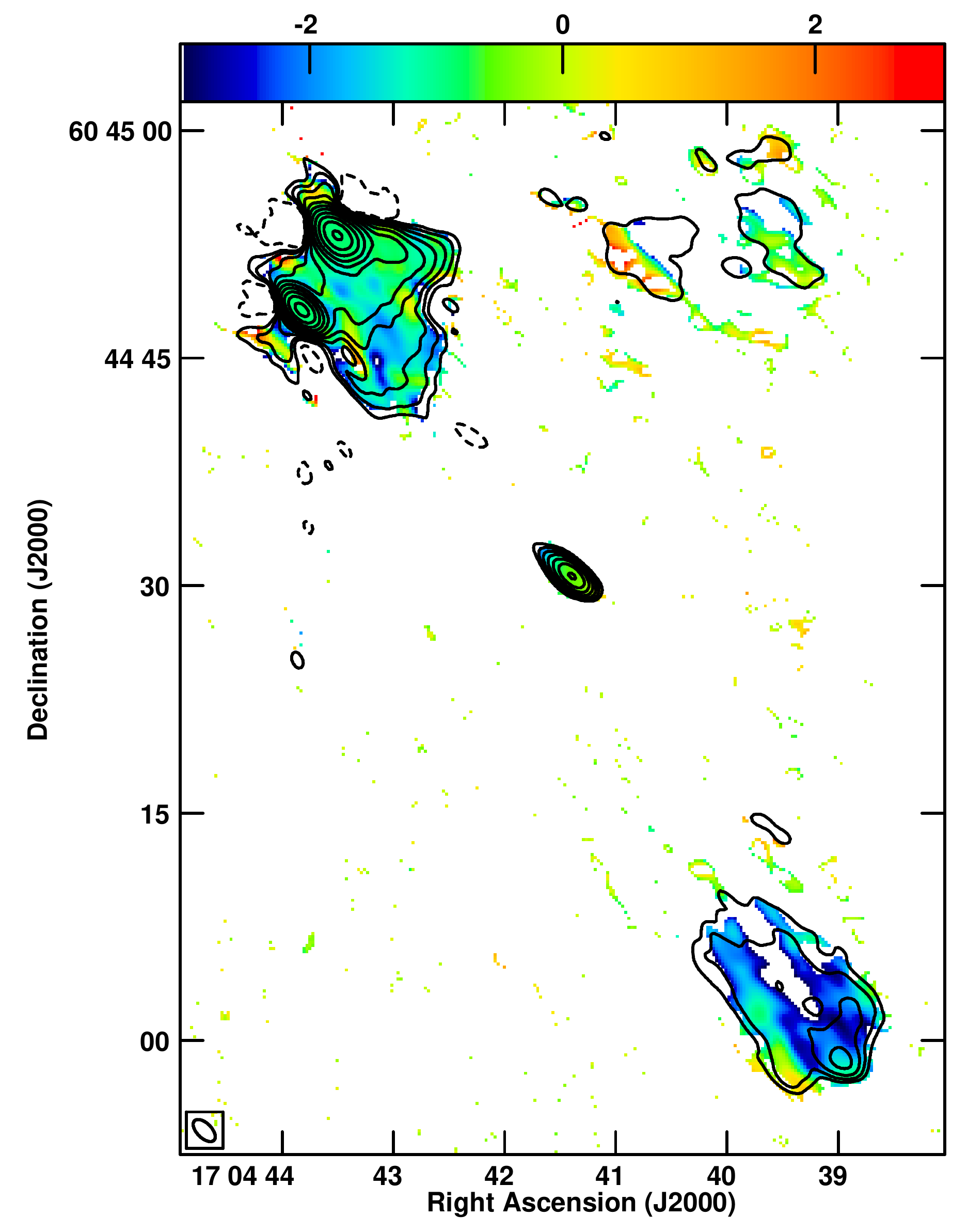}
\caption{\small VLA 6 GHz contour image of quasar PG1704+608 (3C351) superimposed with red (top) polarized intensity vectors and (bottom) in-band spectral index image. The beam is $1.82\arcsec \times 0.97 \arcsec$ with a PA of $44.86^\circ$. The peak surface brightness, $I_P$ is 317.2~mJy~beam$^{-1}$. The contour levels in percentage of the peak surface brightness $I_P$ are $(-0.09,~0.09,~0.18,~0.35,~0.7,~1.4,~2.8,~5.6,~11.25,~22.5,~45,~90)$~Jy~beam$^{-1}$. The length of the EVPA vectors is proportional to polarized intensity with $5\arcsec$ corresponding to 4.16~mJy~beam$^{-1}$.
}
\label{fig9}
\vspace{-28.63836pt} 
\end{figure*}

This SSRQ ({\it aka} 3C351) shows the presence of multiple hotspots and a one-sided distored radio lobe in VLA and GMRT images \citep{Bridle1994, Vaddi2019}. A 15~GHz image from the Cambridge 5-km telescope shows that in the northern hotspot, the southern component is compact while the northern component shows extended emission \citep{Laing1981}. Our 5~GHz image shows the complex radio morphology in total intensity and polarization (Figure~\ref{fig9}). The inferred B-field is parallel to the jet direction for the optically thin radio core. The southern hotspot shows the characteristic B-field morphology of a hotspot (transverse to the jet direction). The jet direction appears to have suddenly changed in this source as is also visible in the misaligned northern radio lobe. In in-band spectral index image (Figure~\ref{fig9}) shows similarly steep spectrum in both the northern as well as southern hotspots. Both lobes show diffuse emission that are not well detected in our higher resolution images. { The VLBA image from \citet{Wang2022} does not resolve the core.}\\ \\

\par 
\indent
Overall, we find that the fractional polarization in the radio cores ranges from 0.8\% to 10\% and in the jets/lobes from 10\% to 40\% (see Table \ref{tab4}). The radio cores primarily display inferred B-fields transverse to the jet direction (assuming optically thick emission) except PG1704+608 where the core emission is optically thin the B-field vectors are aligned along the jet direction. The inferred B-fields are aligned parallel to the local jet directions whenever a jet or a jet knot is observed. The inferred B-field becomes transverse in the compact hotspot regions. However, several hotspots like those in PG1004+130 and PG1100+772, show a more complex B-field structure. We discuss these in greater detail in Section~\ref{Disc} ahead. The B-fields in the radio lobes are almost always aligned with the lobe edges.

\section{Global correlations}
We have looked at various global correlations for the PG ``blazar'' sample using the most recent and reliable estimates of BH masses, accretion rates and SFR values. As far as possible, we have relied on estimates that were obtained uniformly for the various blazar sub-classes. These properties are listed in Table~\ref{tab:PG2}. We have used the Kolmogorov-Smirnov (KS) test for evaluating statistical differences between properties of various populations and the Kendall-$\tau$ test to measure the significance of correlations between different properties. We report { the radio jet kinetic power} $\bar{Q}$ without errors as systematic uncertainties in the estimation methods dominate the statistical uncertainty in the data. $M_{BH}$ and $\dot{M}$ do not have reported errors in \citet{Shangguan_2018,Wu_2009} and \citet{Davis2011} due to the same. The errors in $R_c$ are smaller than the symbol sizes in Figure \ref{corr4}.

\begin{figure}
\centering
\includegraphics[width = 9cm,trim=17 25 0 0]{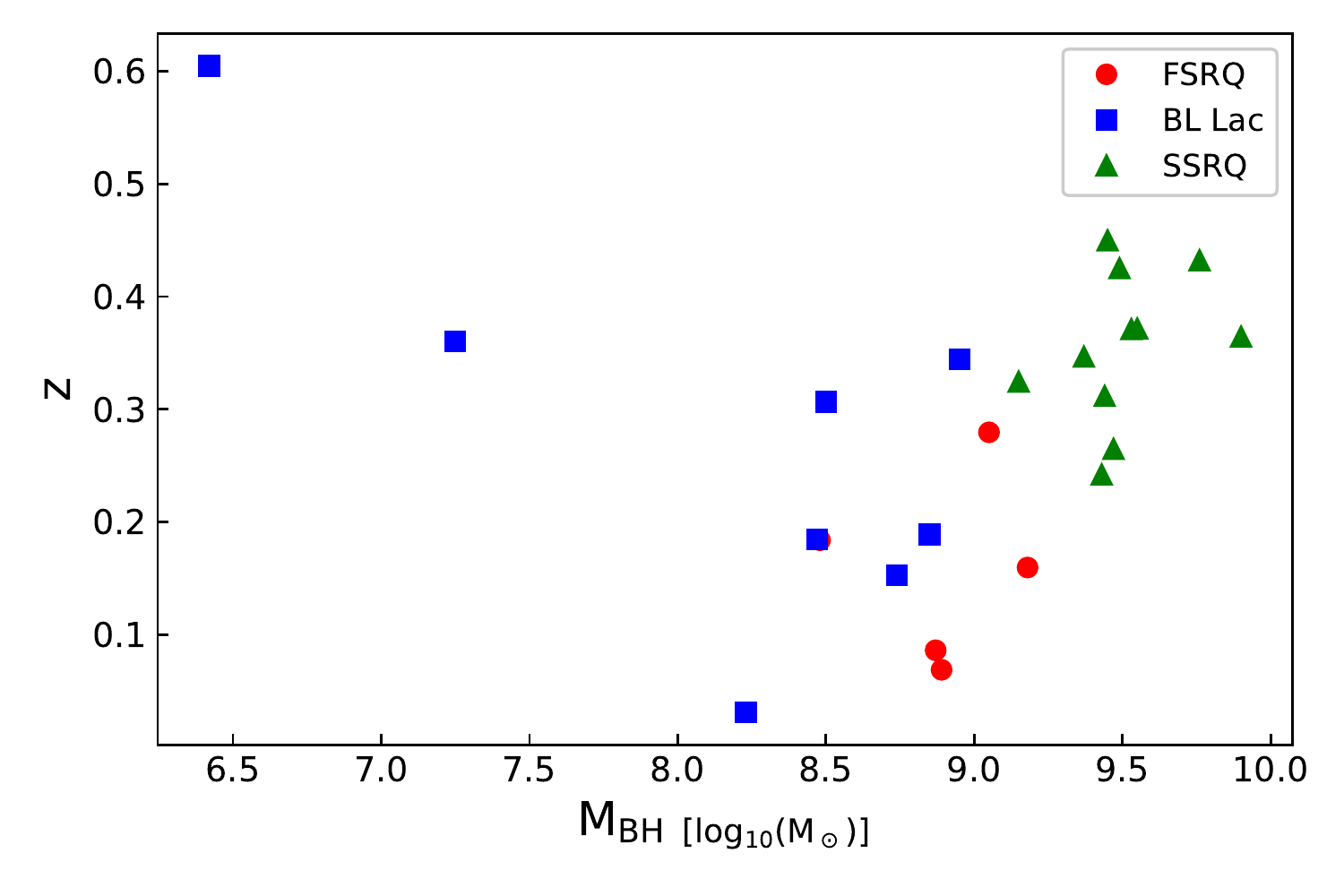}
\caption{\small Distribution of redshifts $z$ w.r.t. the log of BH masses ($M_\mathrm{BH}/M_{\sun}$) for the sample blazars.}
\label{corr2}
\end{figure}

\subsection{Black Hole Mass}
The SMBH masses of the sample blazars were obtained from \citet{Wu_2009} and \citet{Shangguan_2018}. We note that the BH masses were estimated from the H$\beta$ line luminosity in \citet{Shangguan_2018} and from the R-band host galaxy magnitudes in \citet{Wu_2009}; these different mass estimation methods might potentially be introducing biases between the blazar sub-classes. We note that the BH mass estimates obtained by converting a single-band optical luminosity using an empirical correlation, as obtained by \citet{Wu_2009}, may be suspect. This may be the reason for the unusually small BH masses in PG1424+240 and PG1553+113.  

\begin{figure}
\centering
\includegraphics[width = 9.cm,trim=17 25 0 0]{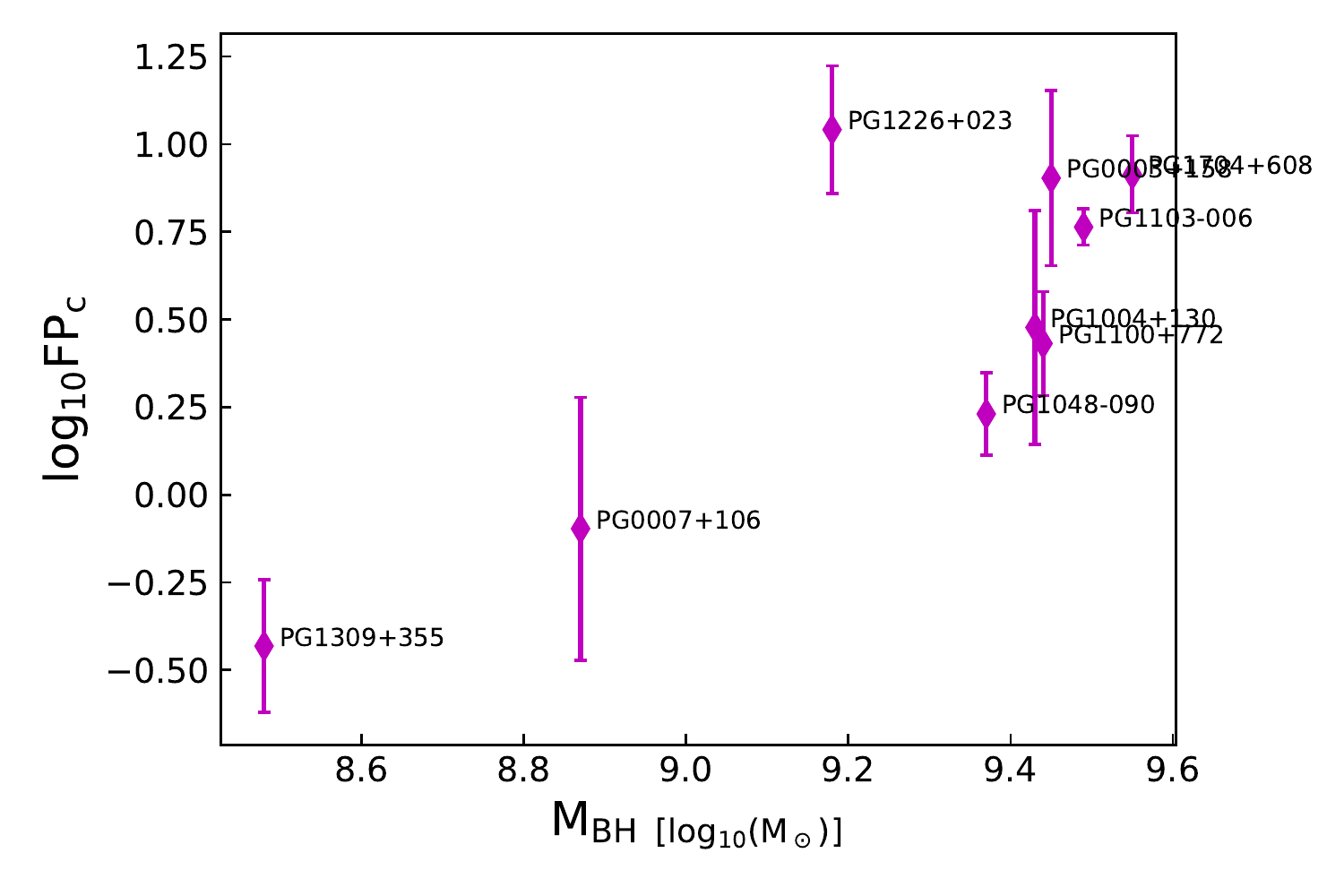}
\caption{\small Log of radio core fractional polarization $\mathrm{FP_C}$ (\%) versus log of BH masses ($M_{BH}/M_{\sun}$) of the quasars.}
\label{corr3}
\end{figure}

\begin{figure}
\centering
\includegraphics[width =9.5cm,trim=15 25 10 0]{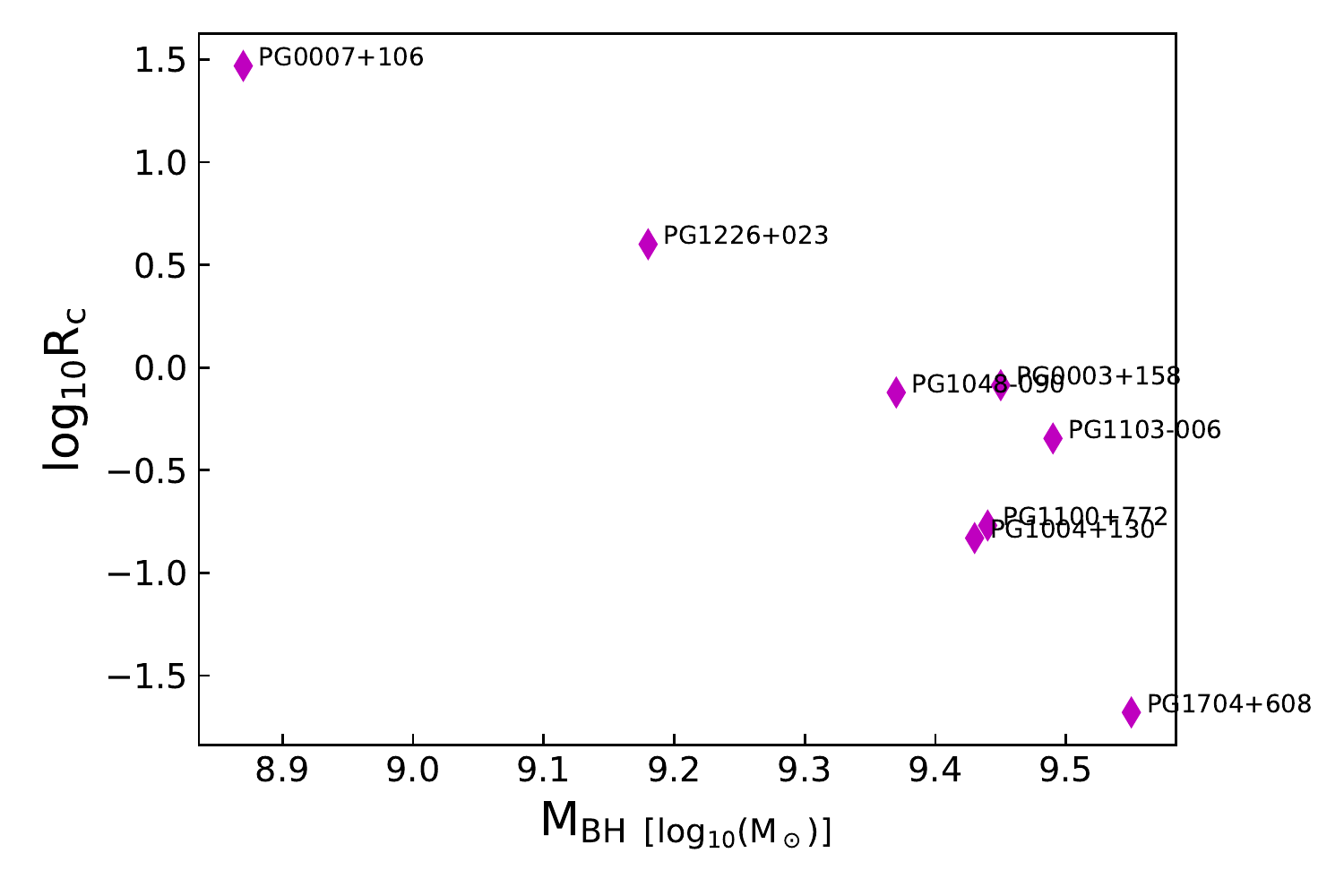}
\caption{\small Log of radio core prominence $R_C$ versus log of BH masses ($M_{BH}/M_{\sun}$) of the quasars.}
\label{corr4}
\end{figure}

The BH masses for the quasars in the current paper however, all come from \citet{Shangguan_2018}. We discuss the possible selection biases affecting the BH masses in the PG sample in Section~\ref{Disc}. The distribution of BH masses $\log_{10} (\mathrm{M_{BH}/M_{\sun}})$ and redshifts for the sample blazars are shown in Figure \ref{corr2}. We find that the hypothesis that quasars and BL~Lacs have different BH masses cannot be ruled out at the 95\% confidence level (KS test p=0.0007). RL quasars have systematically larger BH masses compared to the BL Lacs in our sample. However, we also find that the SSRQs have systematically higher BH masses compared to the FSRQs (KS test p=0.0027), as well as BL~Lacs and FSRQs combined (KS test p=0.00009). So the major BH differences are between the SSRQs and the classical blazars. We find that the BH masses seem to be marginally correlated with the core fractional polarization $\mathrm{FP_C}$ (Kendall $\tau$ test $p=0.044$) and anti-correlated with radio core prominence $R_C$ (Kendall $\tau$ test $p=0.012$) for the 9 quasars studied here (Figures~\ref{corr3} and \ref{corr4}). We discuss the implications of these findings in Section~\ref{Disc}.

\subsection{Star Formation Rate}
The SFR for the PG quasars reported here have been obtained from \citet{Xie2021} except for PG1226+023 which was obtained from \citep{Westhues2016}. The conventional methods to determine SFR for non-active galaxies \citep[which include rest-frame UV-optical SED fitting, optical emission lines, mid infrared (IR) flux density, e.g.,][]{Kennicutt1998,KennicuttEvans2012} may not be reliable for galaxies hosting AGN, especially of Type I. Typically, in the rest-frame far-IR, the contrast between the typical SED of a star forming galaxy and an AGN SED is the largest and the thermal far-IR continuum emission from cool dust grains primarily traces star formation from the host galaxy and is used in most studies \citep[e.g.,][]{Haas2003,Dai2018}. The total IR luminosity (LIR) of the host galaxy was calculated by integrating the \citet{Draine2007} component of the best-fit LIR model. LIR was obtained from 8-1000 $\mu$m measurements. The SFR derived from LIR can only provide us with an upper limit for host galaxy SFR, especially for highly beamed or high luminosity AGN. In the case of BL Lac objects, optical emission from BL Lacs is often dominated by the nuclear source. The FIR and submillimeter also show high flux densities and synchrotron contamination from radio region due to jet triggering and beaming angle \citep{Braun2017}. Therefore, we cannot use LIR as a surrogate for SFR in the case of BL Lac objects. We find that the SFR values are not significantly different between FSRQs and SSRQs (KS test $p=0.27$).

\begin{figure}
\centering
\includegraphics[width =9cm,trim=20 25 0 0]{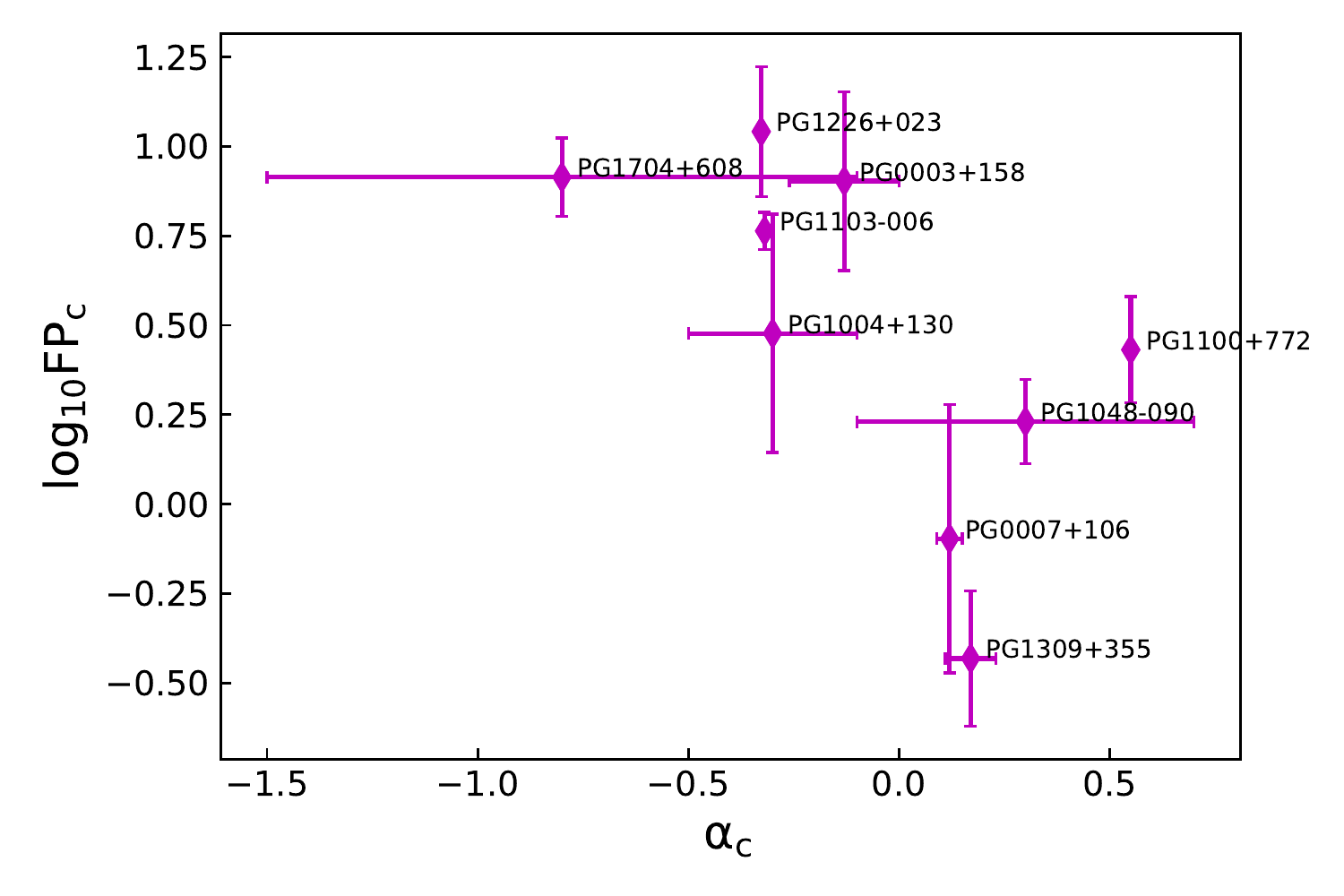}
\caption{\small Log of radio core fractional polarization $\mathrm{FP_C}$ (\%) versus core spectral index $\alpha_C$.}
\label{corr7}
\end{figure}

\begin{figure}
\centering
\includegraphics[width = 9.4cm,trim=20 25 0 0]{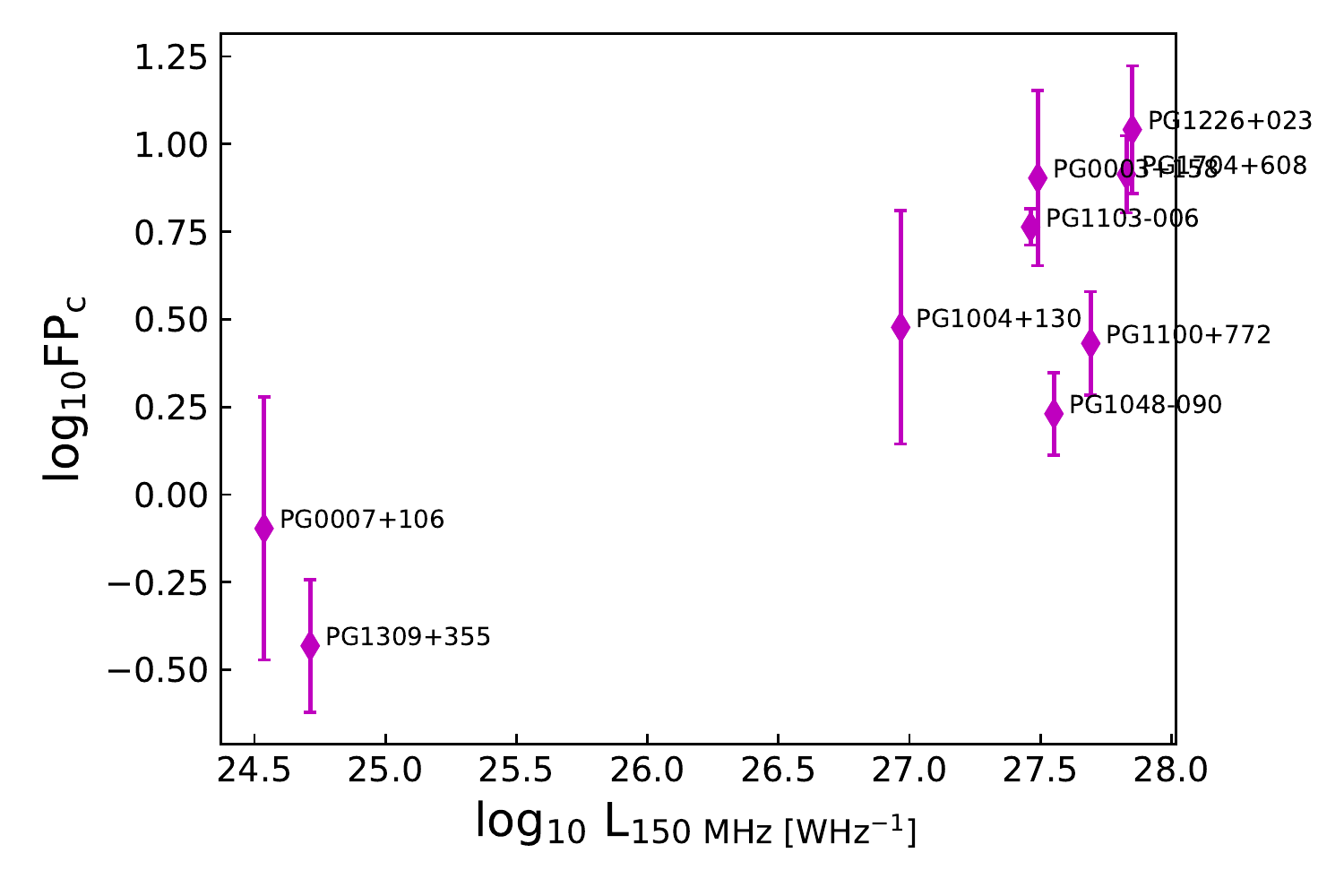}
\caption{\small Log of radio core fractional polarization $\mathrm{FP_C}$ (\%) versus log of total luminosity at 150~MHz ($L_{150~MHz}$) of the quasars.}
\label{corr5}
\end{figure}

\subsection{Jet Power} 
We find a correlation between the radio core fractional polarization and the radio core spectral index (Kendall $\tau$ test $p=0.044$, see Figure~\ref{corr7}) for the 9 quasars; cores with relatively steep spectral indices have a higher fractional polarization. We note that the core spectral indices are typically flat/inverted with the exception of PG1704+608. This implies a relative dominance of unresolved jet emission in the radio core emission.

We find a correlation between the radio core fractional polarization and the 150~MHz total radio luminosity (Kendall $\tau$ test $p=0.024$, see Figure~\ref{corr5}) for the 9 quasars; cores with higher fractional polarization also have higher total radio luminosity at 150~MHz. This could imply that more organized B-fields in the cores lead to higher core fractional polarization and to more radio powerful sources. Interestingly, we found no correlation between the radio core fractional polarization and the accretion rates for the 9 quasars (Kendall $\tau$ test $p=0.180$).

\begin{figure}
\centering
\includegraphics[width = 9cm,trim=15 20 0 0]{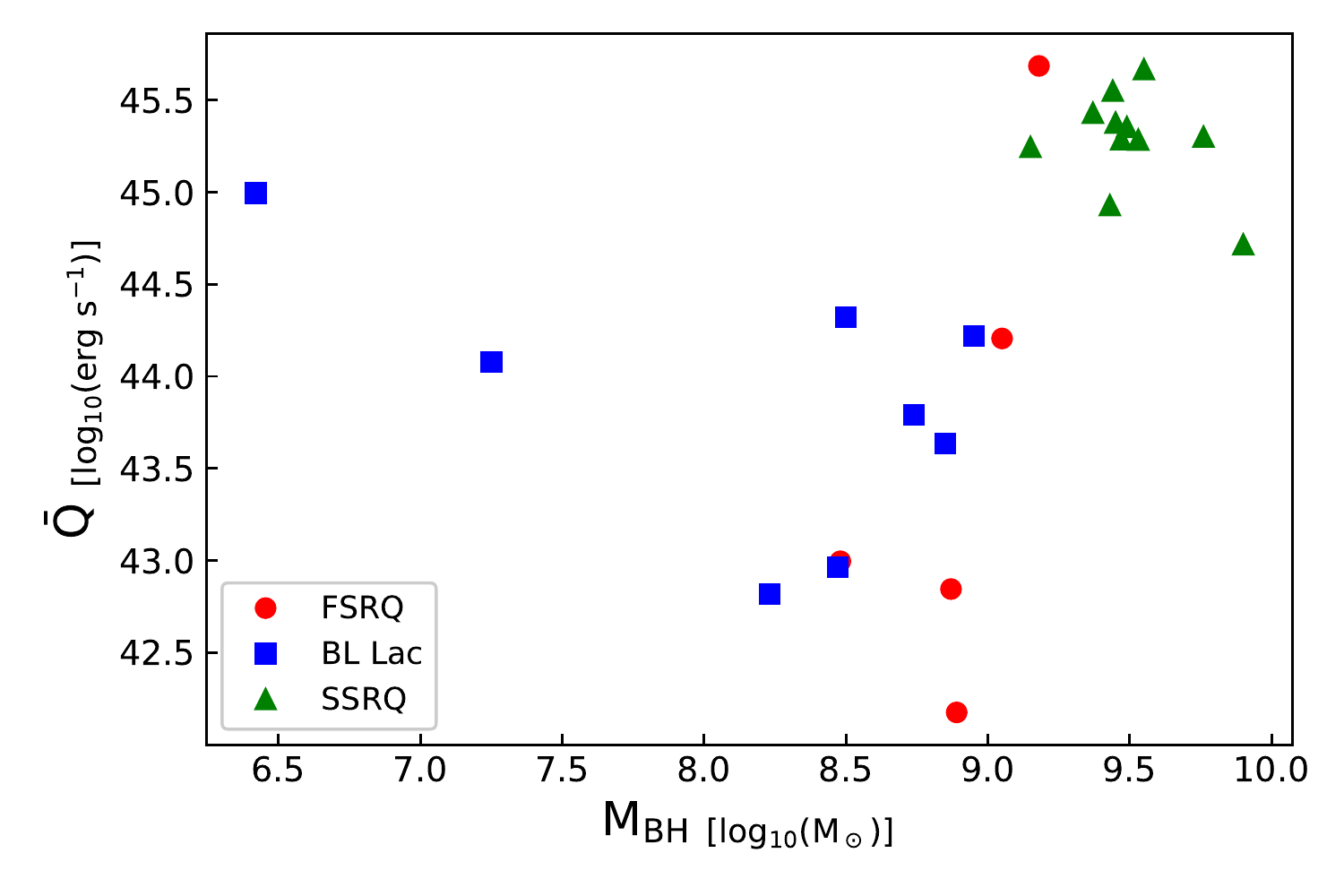}
\caption{\small Distribution of jet power in erg~s$^{-1}$ versus BH masses $\mathrm{\log_{10} (M_{BH} / M_{\sun})}$ for the PG blazar sample.}
\label{corr8}
\end{figure}

\begin{figure}
\centering
\includegraphics[width = 9cm,trim=15 20 0 0]{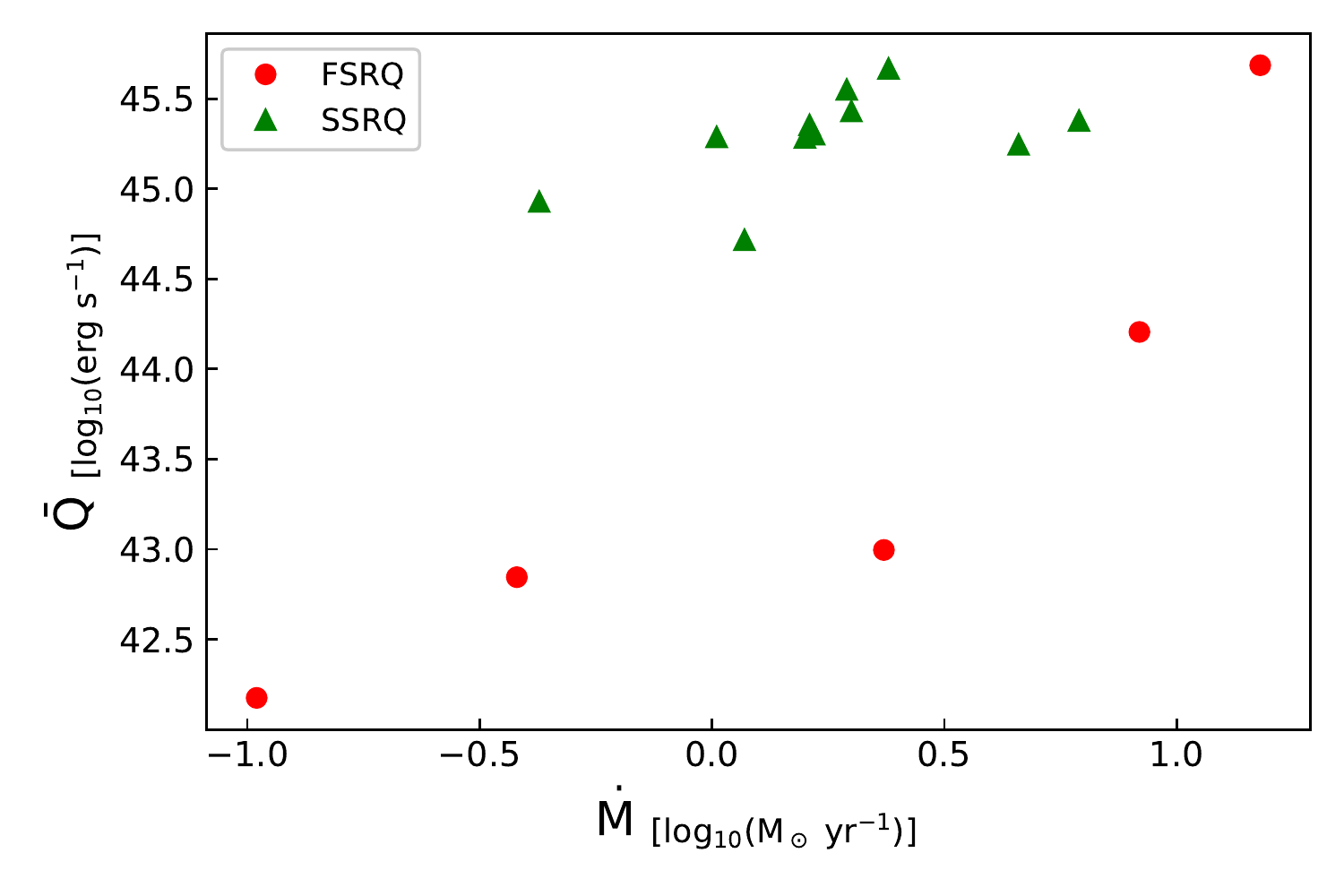}
\caption{\small Distribution of jet power in erg~s$^{-1}$ versus accretion rates $\dot{M}$ for the PG blazar sample.}
\label{corr9}
\end{figure}

\begin{figure}
\centering
\includegraphics[width = 9cm,trim=15 20 0 0]{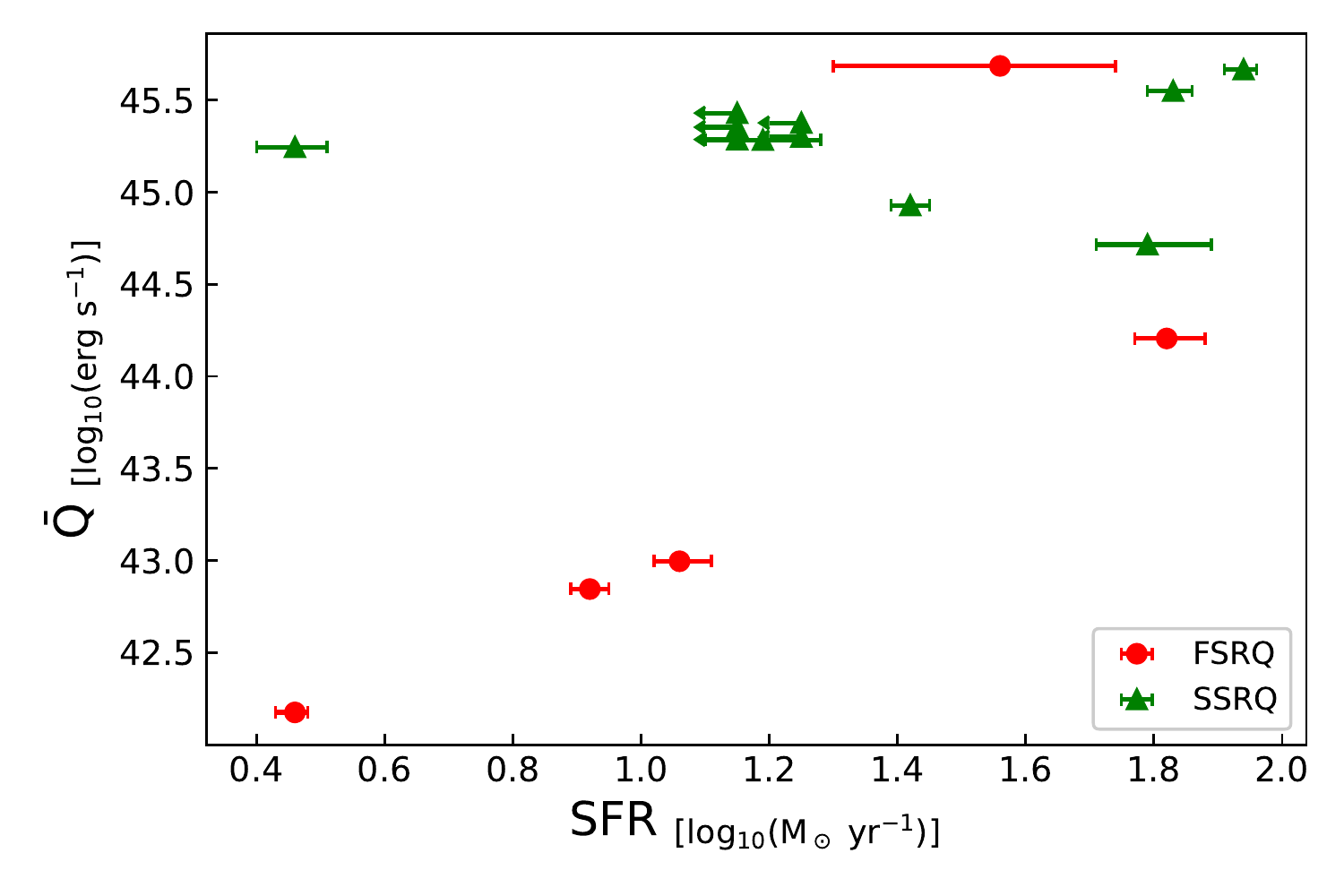}
\caption{\small Distribution of jet power in erg~s$^{-1}$ versus SFR $\mathrm {\log_{10} ( M_{\sun}/yr)}$ for the PG quasars.}
\label{corr10}
\end{figure}

We have estimated the long term, time-averaged bulk radio jet kinetic power $\bar{Q}$ of jets using the radio luminosity at 151 MHz as a surrogate for the luminosity of the radio lobes by using the relation given by \citet{Punsly2018}, 
$$\bar{Q} = 3.8 \times 10^{45} \it{f}~L_{151}^{6/7}~\mathrm{erg~s^{-1}}$$
where $L_{151}$ is radio luminosity at 151~MHz in units of $10^{28}$~W~Hz$^{-1}$~sr$^{-1}$ and $f\approx15$ \citep{Blundell2000}. We obtained the $L_{151}$ from the TGSS survey flux densities \citep{Intema2017} using the relation, $L_{151} = [ D_{L}^2~F_{151}]/[(1+z)^{(1+\alpha)}]$~W~Hz$^{-1}$~sr$^{-1}$, \footnote{We note that there is no $4\pi$ factor because of the anisotropic jet emission, e.g., see \citet{Peacock1999}.} where we have used the typical average $\alpha = -0.7$ for the extended emission. We find that $\bar{Q}$ ranges from $1.50 \times 10^{42}$~erg~s$^{-1}$ to $4.85 \times 10^{45}$~erg~s$^{-1}$ for the PG blazar sample. In Figures~\ref{corr8}-\ref{corr10}, we have examined the relationship between $\bar{Q}$ and BH masses as well as accretion rates. For the entire blazar sample, the jet power appears to be significantly correlated with BH mass (Figure \ref{corr8}; Kendall $\tau$ test $p=0.0025$). However, this correlation disappears when the blazar sub-samples (of FSRQs, SSRQs and BL Lacs) are considered individually, suggesting that there isn't a strong correlation between these two parameters. 

We looked for a correlation between jet power and accretion rates ($\dot M$) for the blazars. We obtained $\dot M$ for the quasars from \citet{Davis2011}, except for the BAL QSO, PG1004+130. For this source, we have adopted an $\eta = 0.1$ value and Eddington ratio $\dot{m_{Edd}}=0.09$ from \citet{Luo2013} and used it to compute its accretion rate. Corresponding $\dot M$ values were not available for the BL~Lacs. We find that the jet power is correlated with accretion rate for both FSRQs (Kendall $\tau$ test $p=0.016$) and SSRQs (Kendall $\tau$ test $p=0.060$) with the former being a stronger correlation. Taken together, the jet power of quasars (FSRQs + SSRQs) remains significantly  correlated with the accretion rate (Figure \ref{corr9}; Kendall $\tau$ test $p=0.020$). Interestingly, we find that the jet power is correlated with the galactic SFR when quasars are considered collectively, (Kendall $\tau$ test $p=0.0515$) but not when taken individually (Figure \ref{corr10}; FSRQ Kendall $\tau$ test $p=0.0833$ and SSRQ Kendall $\tau$ test $p=0.4772$). We discuss the implications of this finding in the next section.

\section{Discussion} \label{Disc}
Our VLA observations reveals that most of the quasars have a disturbed or hybrid FRI/FRII-type radio morphology. Some of them appear to be restarted sources as well, showing clear surface brightness discontinuities between the episodes. We find that in the 9 quasars studied here, the typical hotspot B-field morphology signifying a terminal shock (i.e., B-field perpendicular to the jet direction), is often seen in only one of the hotspots. The inferred B-fields in the radio cores are transverse to the jet direction in all the sources assuming optically thick emission with the exception of PG1704+608 where the B-field are aligned along the jet direction and the core emission is optically thin. The inferred B-fields are aligned parallel to the local jet directions whenever a jet or a jet knot is observed. The in-band spectral index images show that while the radio cores in all sources exhibit flat/inverted spectral indices (with the exception of PG1704+608), the hotspots in several sources have relatively steep spectral indices. This is also consistent with the polarization structures, where the hotspots appear to not be compact terminal shocked regions, but rather locations of jet bends or bow-shock like structures. Such bow-shock like features are consistent with restarted jet activity \citep[e.g.,][]{Clarke1991,Silpa2021}. 

Assuming that the aligned B-fields in the jets signify the dominance of a poloidal B-field component in them, and the transverse B-fields in the cores signify a dominant toroidal B-field component at the unresolved bases of the jets, then we see a transition from a toroidal field to a poloidal field along the radio jets. However, it has been suggested that the toroidal B-field component becomes more dominant and the poloidal component decays faster with distance from the core, since the poloidal field varies as r$^{-2}$ and the toroidal field varies as r$^{-1}$~v$^{-1}$ \citep[e.g.,][]{Begelman1984}. One then needs to invoke some mechanism that can sustain poloidal B-fields over large distances, and explain the quasar results, such as the $\alpha$-turbulence dynamo mechanism \citep{Parker1955}. Alternately, \citet{Silpa2021} had suggested that the transverse to poloidal B-field structure that they had detected in the case of the radio-intermediate quasar III~Zw~2 was an indicator of a `jet+wind' composite radio outflow. It is indeed interesting to note that the majority of the quasars (5/9) would qualify as radio-intermediate quasars \citep[with R values $\lesssim 250$;][see Table~\ref{tab:PG2}]{Falcke1996b}. 

The distorted/hybrid/restarted radio structures could be a consequence of the selection criteria of the PG sample which relies on UV-excess and a star-like nucleus. The PG quasar sample is biased towards luminous sources with high Eddington ratios \citep{Hooper1996,Laor2000,Jester2005}. These properties are closely related to the accretion disk state of the sources and remains largely unbiased in terms of the radio properties of these sources. In principle, this could be suggesting that the kpc-scale radio emission could be influenced by the interaction with its environment. Sensitive low-frequency observations with less restrictive selection effects than earlier studies are also revealing a more complex extended source population, including candidate hybrid radio galaxies, restarting and remnant radio galaxies \citep[e.g.,][]{Kapinska2017,Mingo2019,Jurlin2021}.

The quasar PG1103$-$006 is an X-shaped source. Two primary models have been invoked in the literature to explain such sources: (1) a reorientation of the BH spin axes \citep{Ekers1978,Dennett2002}, (2) superposition of two independent jets produced by two SMBHs residing in the same host galaxy \citep{Lal2007}, and (3) hydrodynamical backflows from the overpressured main jets deflected by the ellipsoidal hot interstellar medium of the host galaxy \citep{Leahy1984,Capetti2002}. The polarization structure observed in PG1103$-$006 is fully consistent with the hydrodynamical backflow model as the inferred B-fields are aligned with the flow of plasma along the X-shaped lobes. It is worth noting that the S-shaped jets are consistent with jet precession, although precession cannot explain the entire structure seen in PG1103$-$006.

\begin{figure}
\centering
\includegraphics[width = 8.8cm, trim=15 25 0 0]{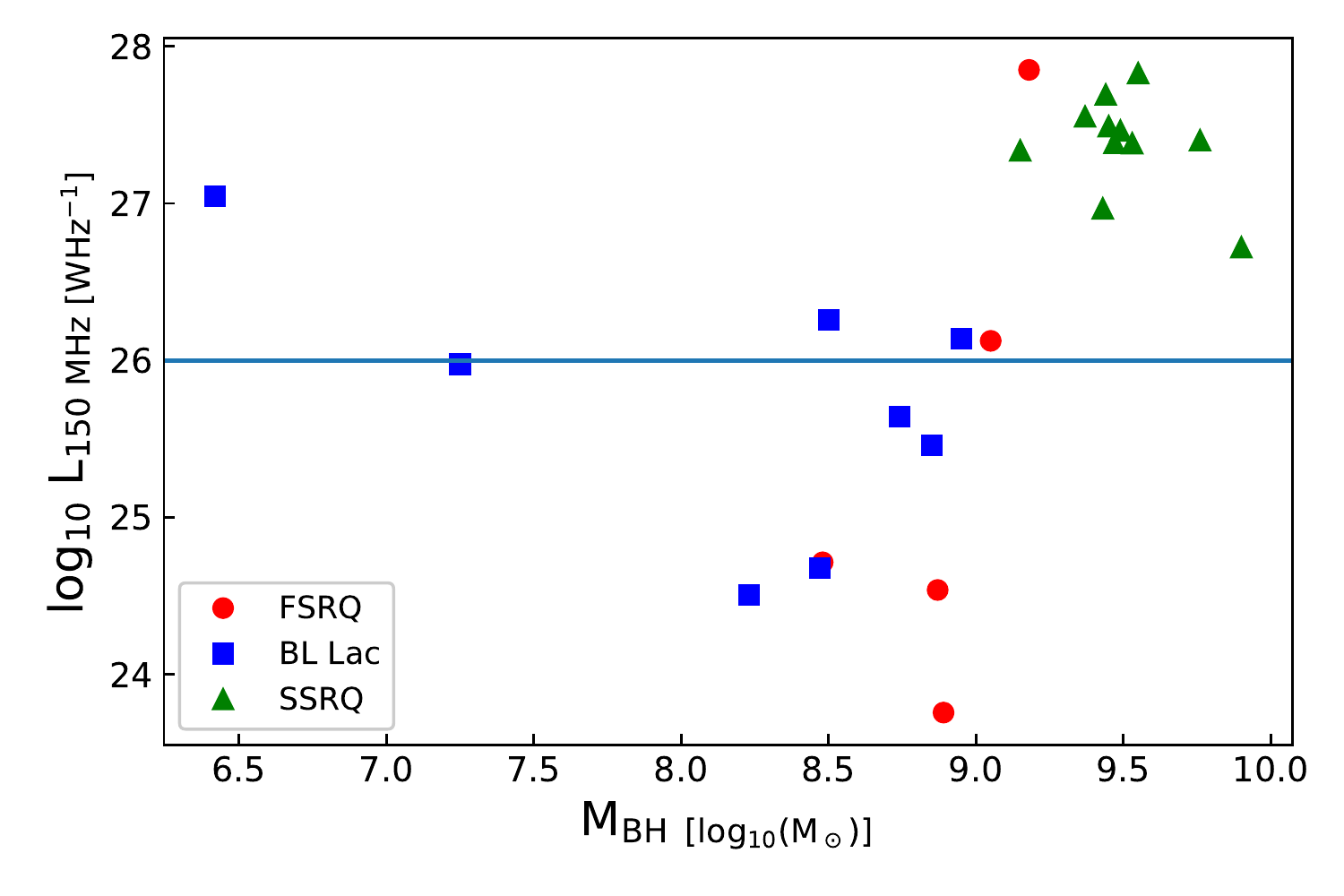}
\caption{\small Distribution of log of total luminosity at 150~MHz ($L_{150~MHz}$) for the PG ``blazar'' sample. The blue line indicates the FRI/FRII luminosity divide derived by \citet{Mingo2019}.}
\label{corr11}
\end{figure}

We have looked at the total radio power at 150~MHz for the sample sources w.r.t. the FR dividing line \citep{Mingo2019} in Figure~\ref{corr11}. {  We have not corrected for Doppler boosting as we did not have reliable Doppler beaming factors for all our sources.} In terms of total radio power at 150~MHz, all the 9 quasars in this paper with the exception of PG0007+106 and PG1309+355 (and PG2209+184 in the larger quasar sample), have FRII-like radio powers. However, not all of them have FRII-like radio morphologies with clear terminal hotspots. It is known that in other samples of radio-loud quasars, not all possess FRII-like (with terminal hotspots) radio morphologies \citep[e.g.,][]{Leahy1991,Black1992,Kharb10}. Three BL~Lac objects, viz., PG0851+203, PG1424+240, and PG1437+398, have FRII-like radio powers. All the remaining BL~Lacs have FRI-like radio powers. Overall, the PG blazar sample sources do not follow the simple unification scheme that suggests that FRI radio galaxies are the parent population of BL~Lac objects and FRII radio galaxies are the parents population of radio-loud quasars. 

We find that PG quasars and BL~Lacs have different BH masses, with the RL quasars having systematically larger BH masses compared to the BL~Lacs. However, we also find that the SSRQs have systematically larger BH masses than the FSRQs, as well as BL~Lacs and FSRQs combined. Therefore, the major difference in BH masses is between SSRQs and the ``classical'' blazars. \citet{Gu2001} have shown that steep- and flat-spectrum quasars have similar distributions of BH masses compared to radio luminosity and radio loudness parameter $R$. However, as noted by \citet{Oshlack2002}, disk-like line emission could change the BH mass estimates by a large factor of 2. Indeed, we find that a factor of 2 change in the BH masses of the FSRQs make them indistinguishable from the BH masses of SSRQs (KS test $p=0.1$).

The optical selection criteria of the PG survey is biased against lower optical luminosity RL quasars \citep{Oshlack2002,Jester2005}. The RL quasars selected on the basis of UV excess tend to be optically luminous with larger BH masses \citep{Oshlack2002} due to the correlation between the host’s bulge mass and the BH mass. We find an anti-correlation between BH masses and radio core prominence $R_C$ for the 9 quasars studied in this paper. This could be a consequence of the line emission originating in a disk \citep[e.g.,][]{Oshlack2002} which would then result in smaller BH masses for larger face-on quasars (see above). BH masses seem to be marginally correlated with the ${\mathrm{FP_C}}$. While this correlation could be suffering from small number statistics, in principle, it could imply a relation between BH mass and more organised/uniform B-fields in jets as they are launched from the BHs. This correlation needs to be further examined with more data. 

The jet power estimates for the PG quasars discussed here come from the calculations of \citet{Punsly2018}. These authors have corrected the \citet{Willott1999} relation for use in blazars by introducing the parameter $\it f$ to account for deviations from 100\% filling factor, minimum energy, a low-frequency cutoff at 10 MHz, the jet axis at $60^\circ$ to the LOS, and no protonic contribution, as well as energy lost expanding the lobe into the external medium, back flow from the head of the lobe, and kinetic turbulence. \citet{Willott1999} had estimated the long term, time-averaged bulk radio jet kinetic power $\bar{Q}$ of jets using the radio luminosity at 151 MHz as a surrogate for the luminosity of the radio lobes, motivated by the assumption that the core emission is attenuated by synchrotron self-absorption at 151 MHz. \citet{Blundell2000} had estimated $\it{f}$ to lie in the range of $10<f<20$ for most FRII radio sources. We used an average value of 15 in this paper. This relation was proposed for FR II radio galaxies and quasars. We adopt this relation to estimate the power of jets in BL Lac objects following \citet{Cao2003} since BL Lac objects have similar radio properties as radio quasars. 

We have found a marginal correlation between the radio core fractional polarization and the radio core spectral index for the 9 quasars; cores with steeper spectral indices have a higher fractional polarization. This implies a dominance of unresolved jet emission in the radio cores of these quasars. 

For the entire blazar sample, the jet power appears to be significantly correlated with BH mass (Kendall $\tau$ test $p=0.0025$). However, this correlation disappears when the blazar sub-samples (of FSRQs, SSRQs and BL Lacs) are considered individually, suggesting that there isn't a strong correlation between these two parameters. The jet power is correlated with accretion rate for both FSRQs (Kendall $\tau$ test $p=0.016$) and SSRQs (Kendall $\tau$ test $p=0.060$) with the former being a stronger correlation. Taken together, the quasars (FSRQs +SSRQs) have their jet power correlated with their accretion rates (Kendall $\tau$ test $p=0.020$). 

We find that the jet power is correlated with the galactic SFR when quasars are considered collectively (Kendall $\tau$ test $p=0.0515$), but not when taken individually (FSRQ Kendall $\tau$ test $p=0.0833$ and SSRQ Kendall $\tau$ test $p=0.4772$). While the marginal correlation between jet powers and the SFR in the host galaxies of the quasars could in principle be consistent with radio-mode feedback, many caveats remain \citep[see][]{Harrison2017,Scholtz2018}. For instance, \citet{ZhuangHo2020,Zhuang2021} have found a correlation between SFR and accretion rates. Additionally, as discussed by \citet{Ward2022}, specific SFR estimates are typically higher in galaxies hosting higher Eddington ratio AGN (for e.g., the PG blazars). Therefore, we cannot draw any conclusions about radio-mode AGN feedback from our current data.

\section{Summary and Conclusions}
In this paper, we have presented results from 6~GHz VLA polarimetric observations of 9 RL quasars. These quasars are a part of our well-selected PG ``blazar'' sample comprising 6 RL quasars and 8 BL Lac objects. We summarise our primary findings below.
\begin{enumerate}
\item The quasars exhibit extensive polarization in their cores, jets/lobes and hotspots, with the exception of PG1309+355, which remains largely unresolved but shows core polarization. The fractional polarization in the cores ranges from 0.8\% to 10\% and jets/lobes from 10\% to 40\%. The inferred B-fields in the radio cores are transverse to the jet direction in all the quasars assuming optically thick emission, with the exception of PG1704+608 where the B-fields are aligned along the jet direction in the optically thin radio core. The inferred B-field directions are largely aligned with the local jet direction in the jets and jet knots of the quasars. The inferred B-fields are transverse to the jet direction in several of the hotspots, consistent with B-field compression in the terminal shock regions. The inferred B-fields in several other hotspots are consistent with jet bends or bow-shock-like structures. Such bow-shock like features are consistent with restarted jet activity. The inferred B-fields in the radio lobes are mostly aligned with the lobe edges. 

\item A large fraction of the quasars show either distorted (S- or X-shaped) morphologies, hybrid FR radio morphologies or restarted activity; the latter is noted as a clear discontinuity in the surface brightness of different features between different episodes. The polarization structures in several hotspots are consistent with restarted jet activity. We attribute this diversity in morphology to the optical/UV selection criteria of the PG sample that remains unbiased at radio frequencies. 

\item The polarization structure observed in the X-shaped quasar PG1103$-$006 is consistent with the hydrodynamical backflow model as the inferred B-fields are aligned with the flow of plasma along the optically thin X-shaped wings. 

\item The in-band spectral index images of these 9 quasars show flat/inverted spectral indices in the radio cores (with the exception of PG1704+608), while the hotspots in several sources have relatively steep spectral indices. This along with the complex polarization structures of the hotspots are also consistent with restarted jet activity in these quasars.

\item For the entire PG ``blazar'' sample (BL~Lacs, FSRQs and SSRQs) we find that they do not fit neatly into the different Fanaroff-Riley total power regime at 150~MHz, as well as Fanaroff-Riley radio morphologies. Several quasar lobes lack a terminal hotspot typical of FRII radio galaxies. Correlations with BH masses appear to be affected by various selection effects related to the mass estimations for the different blazar sub-classes. 

\item We find that the jet mechanical powers correlate with accretion rates for the 16 PG quasars. Accretion rates were not available for the PG BL~Lacs. This correlation implies that higher accretion rates result in more powerful radio jets. We also find a correlation between the radio core fractional polarization and the 150~MHz total radio luminosity for the 9 quasars considered here. That is, cores with higher fractional polarization also have higher total radio luminosity at 150~MHz. This can imply that more organized B-fields at the jet bases lead to higher core fractional polarization and to more radio powerful jets. 
\end{enumerate}

\section*{Acknowledgements}
{  We thank the referee for their constructive suggestions that have significantly improved this manuscript.} JB, PK and SS acknowledge the support of the Department of Atomic Energy, Government of India, under the project 12-R\&D-TFR-5.02-0700. LCH acknowledges and was supported by the National Science Foundation of China (11721303, 11991052, 12011540375) and the China Manned Space Project (CMS-CSST-2021-A04, CMS-CSST-2021-A06). CMH acknowledges funding from an United Kingdom Research and Innovation grant (code: MR/V022830/1). For the purpose of open access, the authors has applied a Creative Commons Attribution (CC BY) licence to any Author Accepted Manuscript version arising. The National Radio Astronomy Observatory is a facility of the National Science Foundation operated under cooperative agreement by Associated Universities, Inc. This research has made use of the NASA/IPAC Extragalactic Database (NED), which is operated by the Jet Propulsion Laboratory, California Institute of Technology, under contract with the National Aeronautics and Space Administration.  This research has made use of the VizieR catalogue access tool, CDS, Strasbourg, France (DOI : 10.26093/cds/vizier). The original description of the VizieR service was published in 2000, A\&AS 143, 23.

\section*{Data Availability}
The data underlying this article will be shared on reasonable request to the corresponding author. The VLA data underlying this article can be obtained from the NRAO Science Data Archive (https://archive.nrao.edu/archive/advquery.jsp) using the proposal ID: 20A-182.

\bibliographystyle{mnras}
\bibliography{main} 

\bsp	
\label{lastpage}
\end{document}